\documentclass[lettersize,journal]{IEEEtran}
\usepackage{amsmath,amsfonts}
\usepackage{algorithmic}
\usepackage{enumitem}
\usepackage[linesnumbered,ruled,vlined]{algorithm2e}
\usepackage{setspace}
\usepackage{array}
\usepackage{textcomp}
\usepackage{stfloats}
\usepackage{multirow,diagbox,tabularx,blindtext}
\usepackage{booktabs}
\usepackage{hyperref}
\usepackage{verbatim}
\usepackage{graphicx}
\usepackage{amssymb}
\usepackage{textcomp}
\usepackage{cite}
\usepackage{color}
\usepackage{pifont}
\usepackage{makecell}
\usepackage{threeparttable}
\usepackage{arydshln}
\usepackage[T1]{fontenc}
\usepackage[scaled=0.81]{beramono}

\newcommand{\PP}[1]{
\vspace{2px}
\noindent{\bf \IfEndWith{#1}{.}{#1}{#1.}}
}

\usepackage[normal]{caption}
\usepackage{array}
\usepackage{arydshln}
\usepackage{subcaption}
\usepackage{enumitem}
\usepackage{colortbl}
\usepackage{tabularx}
\usepackage[dvipsnames]{xcolor}
\usepackage{hyperref}
\definecolor{bgreen}{RGB}{0,170,0}
\definecolor{bred}{RGB}{220,0,0}
\definecolor{mydarkblue}{RGB}{0,0,150}
\definecolor{Gray}{gray}{0.93}
\hypersetup{
    colorlinks=true,
    linkcolor=bred,
    citecolor=mydarkblue,
    filecolor=bred,
    urlcolor=mydarkblue
}
\newcommand{\todo}[1]{}
\renewcommand{\todo}[1]{{\color{red} TODO: {#1}}}

\begin{document}

\begin{sloppypar}

\title{\huge Bridging Source Code and Bytecode for Smart Contract Vulnerability Detection via Dual-Perspective Cross-Modal Distillation} 

\author{Ye Tian, Yifan Jia, Yanbin Wang*, Jianguo Sun*, Haitao Xu, Xin Wang, Zhihua Fu

\thanks{\emph{* Corresponding author: Yanbin Wang and Jianguo Sun.} }       

\thanks{

Ye Tian, Jianguo Sun, Xin Wang and Zhihua Fu are with the Hangzhou Research Institute, Xidian University, Hangzhou, 311231, China.
Yifan Jia is with the Harbin Engineering University, Harbin, 150001, China. 
Yanbin Wang is with the Shenzhen MSU-BIT University, Shenzhen, 518172, China and School of Computer Science and Technology, Beijing Institute of Technology, Beijing, 100081, China. 
Haitao Xu is with the College of Computer Science, Zhejiang University, Hangzhou 310058, China. }
}

\markboth{IEEE Transactions on Dependable and Secure Computing} 
{Ye Tian \MakeLowercase{\textit{et al.}}:Bridging Source Code and Bytecode for Smart Contract Vulnerability Detection via Dual-Perspective Cross-Modal Distillation} 


\maketitle

\begin{abstract}
Smart contract vulnerabilities have caused substantial financial losses, yet most deployed contracts are closed-source, forcing detection to operate on bytecode --- which lacks the high-level semantic information available in source code. To compensate, recent cross-modal methods distill knowledge from source-code models into bytecode detectors by aligning the two modalities. However, these methods align source and bytecode only at the graph level through global embedding matching, whereas vulnerability is a property of specific nodes and control-flow regions --- so the student preserves global structure but loses the fine-grained node-level correspondences that decide a contract's safety. We propose ExDoS, a dual-focus cross-modal distillation framework that addresses this limitation from three angles. First, We introduce aligned vulnerability patterns that mark corresponding nodes in both source and bytecode graphs, establishing the missing node-level supervision and cross-modal correspondence; with these correspondences in place, we then propose a dual-attention graph network that applies relation-aware attention and adaptive node weighting so that vulnerable nodes survive into the graph-level embedding rather than being diluted by uniform aggregation; given preserved node-level signals and established correspondences, we also propose a dual-focus distillation objective whose global loss retains whole-graph alignment while its local loss matches expert-paired nodes to enforce region-level consistency. On real-world contracts, ExDoS reaches F1 of 90.86\%, 90.23\%, and 83.94\% for reentrancy, timestamp dependency, and infinite loop --- improving by 2.7--5.1 points over the strongest per-type baseline. Our ablations confirm that the pattern annotations, local alignment, and attentive encoding each address a distinct limitation of prior distillation-based approaches.
\end{abstract}


\section{Introduction}
 Blockchain technology has enabled decentralized, immutable, and transparent systems, fostering applications across decentralized finance, supply chain management, and digital asset exchanges. At the core of many such applications are smart contracts---self-executing programs that automate transactions without centralized intermediaries. Despite their advantages, smart contracts have shown significant susceptibility to security vulnerabilities \cite{chen2025numscout,zheng2024dappscan}. Exploitable flaws cause not only substantial financial losses \cite{chen2025chatgpt} but also disruptions to the operational integrity of decentralized systems; for instance, the 2023 Euler Finance exploit resulted in approximately \$200 million in losses \cite{chainalysis2023euler}. Such incidents undermine the foundational promise of trustless agreements and weaken confidence in the broader ecosystem.

A broad spectrum of techniques has been developed for smart-contract vulnerability detection, including formal verification \cite{chen2024intro_formal3}, symbolic execution \cite{so2021intro_symbolicexe1,yao2022introsymbolicexe2}, and fuzz testing \cite{he2019intro_fuzz1,nguyen2020sfuzz,ye2024funfuzz}. While effective for known patterns, these rule-centric approaches demand substantial expert effort, are tightly coupled to source code, and become costly to maintain at scale \cite{wu2024handle_expert_challenge,qian2022handle_expert_challenge,ding2025LLM4scv_handle_expert_challenge}. To improve scalability and generalization, data-driven learning on contract artifacts has gained traction \cite{hu2021dl_lstm,liu2022dl_GraphTransformer,duan2023new}. Many such methods operate on source code and benefit from its rich semantics; however, in realistic deployment, most contracts are closed-source and only bytecode is accessible. Recent cross-modal distillation methods \cite{sun2025mtvhunter,qian2023cross_www} transfer a source-code teacher's knowledge to a bytecode-only student, compensating for the semantics lost when source is unavailable---an important step toward practical detection.

This transfer, however, rests on an implicit assumption: that aligning the two modalities at the graph level is sufficient to propagate vulnerability-relevant information. Vulnerability is a property of specific nodes and control-flow regions, not of the contract as a whole. When alignment operates only at the graph level, three consecutive failures undermine the transfer:
\begin{itemize}
\item \textbf{F1 -- missing node-level supervision.} Although the source side carries expert-defined vulnerability patterns and node-level annotations, the bytecode side carries none, leaving the distillation objective without the local signal needed to decide which bytecode regions match the flawed source nodes.
\item \textbf{F2 -- loss of node-level signal in encoding.} Conventional graph encoders aggregate nodes and edges uniformly without distinguishing vulnerability-carrying structure, so the few vulnerability-critical nodes are averaged into the surrounding normal ones; even with annotations in place, the objective then receives representations that have already lost the local signal.
\item \textbf{F3 -- global-only alignment objective.} The distillation objective matches only global embeddings, with no loss term enforcing regional consistency where vulnerabilities actually manifest; the student therefore treats each contract as a single whole and remains blind to the specific nodes that decide its safety.
\end{itemize}
These failures are not independent: without node-level annotations a local alignment term cannot even be formulated (F1); without attentive encoding that preserves vulnerable nodes, such a term would operate on degraded inputs (F2); and without a local objective, global alignment alone cannot close the node-level gap (F3).

We propose ExDoS, a dual-focus cross-modal distillation framework that addresses this chained failure from three coordinated angles. We introduce cross-modal vulnerability pattern annotations that mark corresponding nodes in both source and bytecode graphs, establishing the missing node-level supervision and the correspondence dictionary needed for local alignment. With these correspondences in place, we develop a dual-attention graph network that applies relation-aware attention and adaptive node weighting so that vulnerable nodes survive into the graph-level embedding rather than being diluted by uniform aggregation. Given preserved node-level signals and established correspondences, we formulate a dual-focus distillation objective whose global loss retains whole-graph alignment while its local loss matches expert-paired nodes to enforce regional consistency. After distillation, the student operates on bytecode alone, requiring no source code at inference.

We evaluate ExDoS on real-world contracts across reentrancy, timestamp dependency, and infinite loop vulnerabilities. Our contributions are summarized as follows:
\begin{itemize}
\item We propose ExDoS, which advances bytecode-based vulnerability detection by transferring source-code knowledge to bytecode representations prior to deployment---without requiring source code at inference---and improves F1-scores by 2.7--5.1 points over the strongest per-type baseline across reentrancy, timestamp dependence, and infinite loops.

\item We develop the Dual-Attention Graph Network (DAGN), which introduces relation-aware attention and learnable node attention pooling to complement graph attention. This dual design preserves the vulnerability-critical node signals that uniform encoding would otherwise dilute: relation-aware attention differentiates vulnerability-carrying edges, while adaptive pooling lets the few critical nodes survive aggregation rather than being averaged into the majority.

\item We introduce the first aligned bytecode-level vulnerability patterns for the target vulnerabilities by summarizing existing source-code patterns and designing corresponding opcode-level counterparts. The resulting expert annotations provide fine-grained supervision at the block/node level and guide the model toward vulnerability-critical regions during training.

\item We formulate a dual-focus distillation objective comprising a global semantic loss that aligns graph-level embeddings for holistic transfer, and a local structural loss that aligns expert-annotated nodes to establish vulnerability-wise cross-modal consistency.
\end{itemize}

\section{Backgroud and Related work}

\label{sec:backgroud}
\subsection{Blockchain and Smart Contracts}

Blockchain is a decentralized, append-only ledger that enables secure and verifiable transactions among untrusted parties \cite{nakamoto2008bitcoin}. Transactions are grouped into cryptographically linked blocks and distributed across network nodes to ensure integrity and transparency. Among blockchain platforms, Ethereum is widely adopted for its support of smart contracts—self-executing programs that enforce agreements automatically \cite{buterin2013ethereum,lin2024crpwarner,guo2024smart}.

Smart contracts are typically written in languages such as Solidity and compiled into bytecode for execution on the Ethereum Virtual Machine (EVM). Once deployed, they operate deterministically without further intervention. However, their immutability means that any vulnerabilities or logic flaws can lead to irreversible consequences. From a security perspective, smart contracts face several challenges. They run in adversarial, permissionless environments where minor flaws can be exploited for financial gain. Their complexity—arising from inter-contract calls, asynchronous execution, and persistent state—hinders analysis. Moreover, many contracts lack publicly available source code, requiring analysis at the bytecode level with limited semantic information \cite{chen2020bytecodeyes_sccodeno_argue}. These issues call for robust and scalable vulnerability detection methods applicable even without source code.

\subsection{Vulnerabilities Description}
Numerous studies show that smart contracts are prone to various security vulnerabilities caused by programming errors, incomplete specifications, and platform-specific semantics \cite{singh2020bkg2.2_r2,mense2018bkg2.2_r3/timestamp}. Among them, reentrancy, timestamp dependency, and infinite loop execution are particularly prevalent and damaging \cite{chen2025clep,wu2024bkg2.2_r4_loop}. We briefly describe them below.

\subsubsection{Reentrancy}

\begin{figure}[t]
    \centering     
\includegraphics[width=0.99\linewidth]{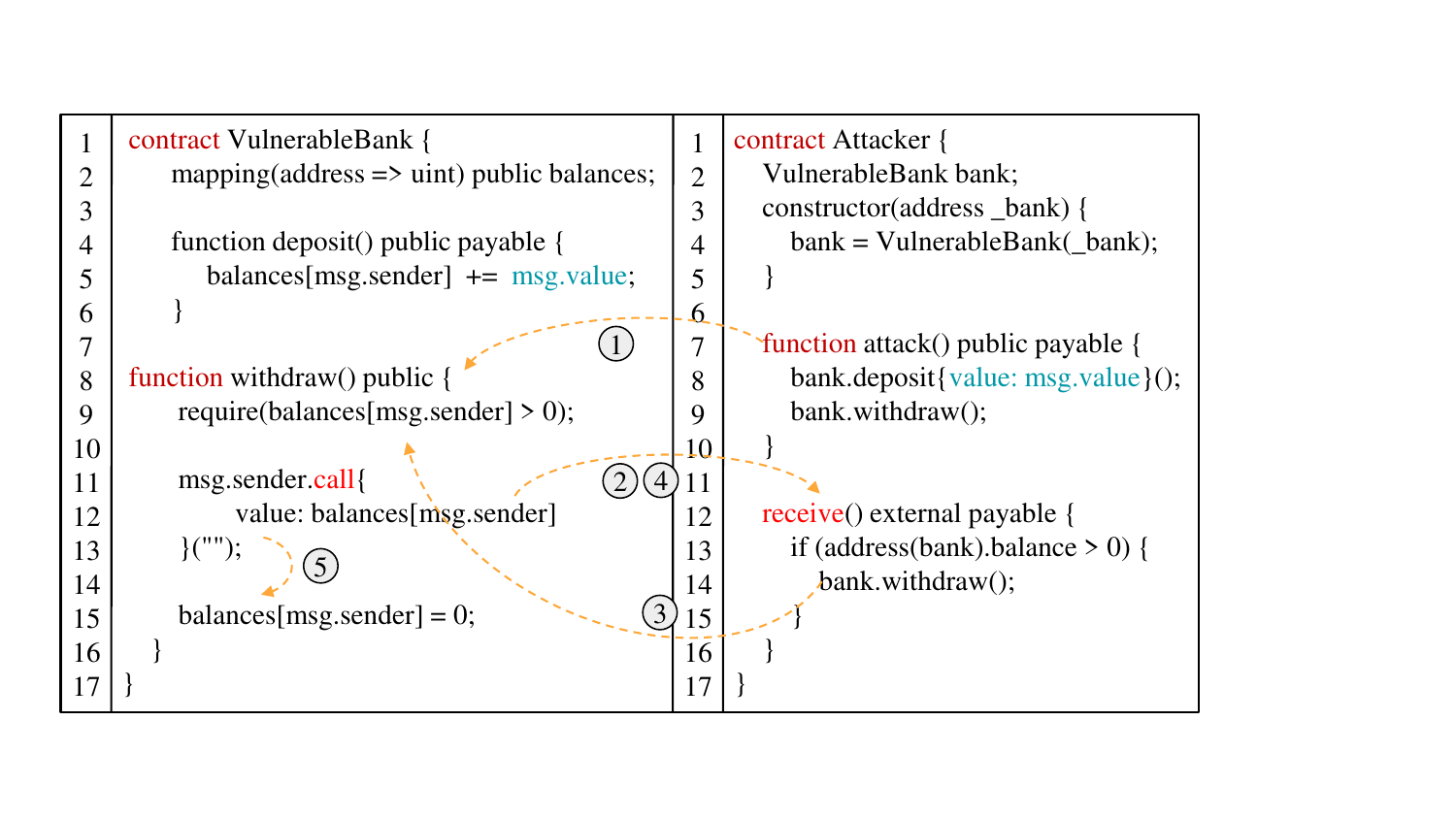}
    \caption{The vulnerable contract VulnerableBank violates the checks-effects-interactions pattern. The attacker re-enters the withdraw function before state updates, enabling repeated withdrawals.}
    \label{fig:reentrancy example}
\end{figure}

Reentrancy occurs when an external contract re-enters a function before its execution completes, disrupting control flow and state consistency. A classic example is the 2016 DAO attack, where recursive withdrawals led to losses exceeding \$60 million \cite{zhao2017dao}. This issue often arises when external calls are made before updating internal state, violating the checks-effects-interactions pattern. In Figure \ref{fig:reentrancy example}, the contract transfers Ether before updating balances, allowing an attacker to repeatedly invoke the withdraw function and drain funds.

\subsubsection{Timestamp Dependency}
Timestamp dependency arises when contracts rely on block timestamps for critical decisions. Since miners can slightly manipulate timestamps, attackers may exploit this to influence outcomes such as lotteries or time-based logic. Although commonly used as time proxies, timestamps are nondeterministic and unsuitable for security-sensitive operations \cite{mense2018bkg2.2_r3/timestamp}.

\subsubsection{Infinite Loop}
Smart contracts are constrained by gas limits, which bound computational resources. Unbounded or input-dependent loops may exceed this limit, causing execution failure. Malicious contracts can exploit this to trigger denial-of-service (DoS) conditions in dependent contracts \cite{wu2024bkg2.2_r4_loop}. Detecting such vulnerabilities is difficult, as they often depend on complex control flow and inter-contract interactions.

These vulnerabilities represent critical failure modes that can cause significant financial and systemic damage, as confirmed by empirical studies and real-world incidents \cite{wu2024handle_expert_challenge}. Therefore, automated detection techniques are essential for improving blockchain security.

\subsection{Smart Contract Vulnerability Detection}

Early detection relied on program analysis tailored to contract artifacts. Formal-methods tools such as Securify and Teether model and verify contract properties through assertions \cite{tsankov2018securify,krupp2018teether}, while symbolic-execution tools such as Oyente and the Mythril framework explore the control-flow graph to surface reentrancy, integer overflow, and assertion failures \cite{luu2016making,mueller2017framework}; fuzzing complements them by generating and mutating transaction sequences to maximize coverage \cite{nguyen2020sfuzz,liu2023rethinking}. These techniques offer precision on known patterns, but they depend on manually defined specifications, incur high analysis cost at scale, and generalize poorly to unseen contract structures.

Learning-based detection removes the specification burden by training on contract representations. Early work applied classical classifiers to analyzer outputs \cite{momeni2019machine} or sequential models to bytecode \cite{tann2018towards}. Graph-based methods then became a prevailing representation, constructing syntactic or semantic contract graphs and applying graph neural networks \cite{zhuang2021smart_tmp,luo2024scvhunter}, sometimes hybridizing expert vulnerability patterns with learned representations \cite{liu2021combining_expert_gnn,liu2021interpretable_graph_expert_ame}. Source-dependent graph models capture rich semantics but presuppose source availability, which most deployed contracts do not provide. To bridge this gap, cross-modal distillation transfers a source-code teacher's knowledge to a bytecode-only student \cite{qian2023cross_www,sun2025mtvhunter,chen2024contrastive_ICSE}; this is a necessary step toward closed-source detection, yet it aligns the two modalities only at the graph level through global embedding matching and uniform node aggregation, leaving the student without the node-level correspondences that decide a contract's safety and diluting the few vulnerable nodes into the majority.

Complementing these source-dependent or cross-modal approaches, a separate line of work detects vulnerabilities directly from bytecode without requiring source code. Chen et al. proposed DefectChecker, which mines expert features from EVM bytecode for defect detection \cite{chen2021tradition_byteonly}. Shi et al. classified contracts from raw bytecode via structural learning \cite{shi2022byteonly}, while Li et al. introduced VulHunter, a multiple-instance-learning detector operating at the EVM bytecode level \cite{li2023vulhunter}, and Cobra, an interaction-aware bytecode-level detector \cite{li2024cobra}. Bu et al. further proposed SmartBugBERT, a BERT-based bytecode detector \cite{bu2025smartbugbert_byteonly}. These methods target the same closed-source deployment setting as our work but rely solely on bytecode cues, whereas ExDoS transfers source-code knowledge to a bytecode-only student so that the student still benefits from source semantics at training time.


\section{Method}
\begin{figure*}[t]
    \centering     
\includegraphics[width=0.92\linewidth]{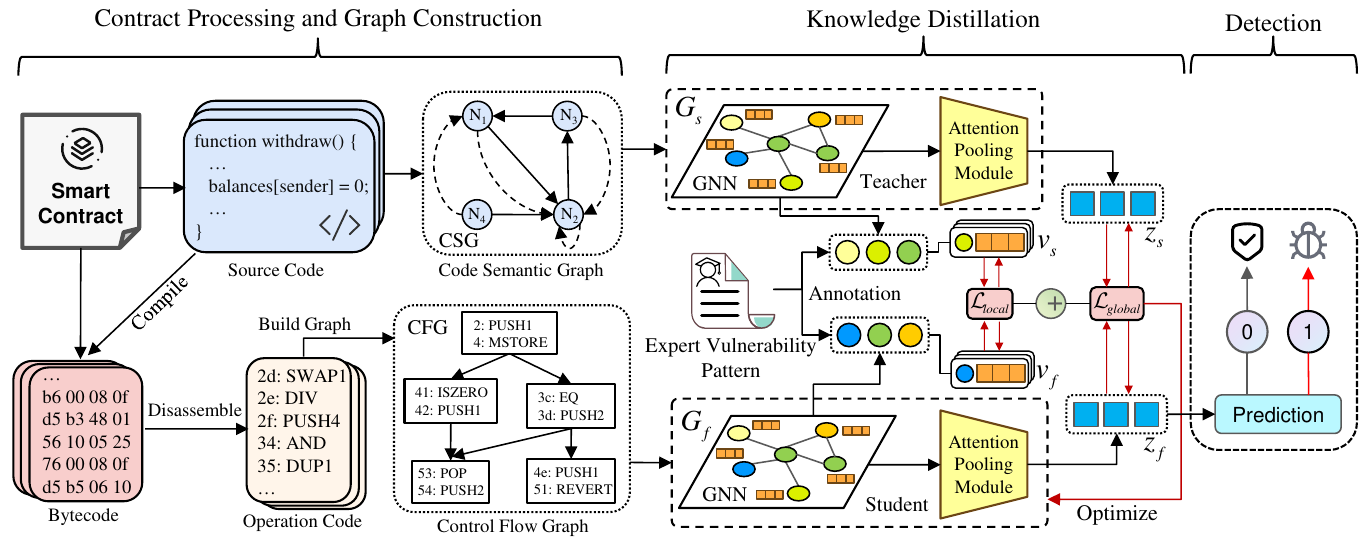}
    \caption{The framework of Proposed Expert-Guided ExDoS Framework.}
    \label{fig:framework}
\end{figure*}

This section outlines the overall design of our proposed ExDoS Framework. Figure \ref{fig:framework} provides a high-level overview of our proposed pipeline.

Cross-modal distillation transfers knowledge from a source-code teacher to a bytecode-only student by aligning their graph representations. This transfer, however, rests on an implicit assumption: that aligning at the graph level is sufficient to propagate vulnerability-relevant information across modalities. In practice the assumption fails along three consecutive links of the information chain. First, bytecode carries no formalized vulnerability patterns or node-level annotations, so the distillation objective lacks the local supervision needed to identify which regions should be aligned. Second, even when such annotations exist, conventional graph encoders treat nodes and edges uniformly---aggregating without distinguishing vulnerability-carrying structure and pooling homogeneously---so the few vulnerability-critical nodes neither stand out in their embeddings nor survive into the graph-level representation, and the alignment objective operates on representations that have already lost the local signal. Third, existing distillation objectives optimize only global embedding proximity, offering no mechanism to enforce consistency at the region where vulnerabilities actually manifest. These failures are not independent: without annotated correspondences a local alignment term cannot even be formulated; without attentive encoding that preserves vulnerable nodes, such a term would operate on degraded inputs; and without a local objective, global alignment alone cannot close the node-level gap. We address this chained failure through three coordinated mechanisms---an aligned pattern annotation scheme, a dual-attention graph network that preserves vulnerability-critical nodes, and a dual-focus distillation objective---and present each in turn.

During training, we use a labeled dataset of smart contracts, each with both source code and corresponding bytecode. We transform each modality into a graph representation---a Code Semantic Graph (CSG) from source code and a Control Flow Graph (CFG) from bytecode---capturing control-flow and data-flow dependencies. To encode these graphs while preserving vulnerability-critical nodes, we apply a Dual-Attention Graph Network (DAGN) that combines relation-aware attention with adaptive node weighting. We further introduce expert-guided vulnerability patterns annotated on both graphs, establishing the node-level correspondences the distillation objective requires. Building on this, we design a cross-modal student--teacher distillation framework in which a teacher pretrained on source CSGs transfers knowledge to a bytecode-only student through a dual-focus objective: a global loss aligns graph-level embeddings, while a local loss matches expert-paired nodes to enforce region-level consistency. At inference, the teacher is discarded and the student detects vulnerabilities directly from bytecode. The following subsections detail each component.

\subsection{Contract Graph Construction and Encoding}

Graph-based representations are widely adopted in program analysis due to their ability to capture rich control-flow and data-flow semantics \cite{luo2024scvhunter,sun2025mtvhunter}. In our framework, we construct two modality-specific graph representations for each smart contract: a \textit{Code Semantic Graph} (CSG), denoted as $G_s = (V_s, E_s)$, derived from source code, and a \textit{Control Flow Graph} (CFG), denoted as $G_f = (V_f, E_f)$, extracted from bytecode. These graphs serve as structural foundations for subsequent representation learning.

\subsubsection{Source Code (CSG)}
For smart contracts with accessible source code, we construct the Code Semantic Graph to encode the program’s internal semantic relationships. Each node $v_i \in V_s$ represents a program element such as a variable, function call, or control structure, while each directed edge $e_{ij} \in E_s$ captures either control-flow or data-flow dependency. Edges are temporally ordered to preserve the execution sequence as it appears in the original source code, enabling a faithful representation of the contract's operational logic.


This graph-based abstraction facilitates the integration of structural and semantic features into downstream learning by capturing both control-flow and data-flow relations, thereby providing a unified representation that enhances the expressiveness of contract embeddings.

\subsubsection{Bytecode (CFG)}
Smart contract bytecode is typically represented as hexadecimal sequences, which are difficult to interpret directly. To recover higher-level control structures, we first apply an automated tool $BinaryCFGExtractor$\footnote{https://github.com/Messi-Q/BinaryCFGExtractor} to translate bytecode into opcode sequences. These opcodes provide a more interpretable abstraction, exposing the underlying control and data semantics embedded in the compiled binary.

\begin{figure*}[t]
    \centering     
\includegraphics[width=1\linewidth]{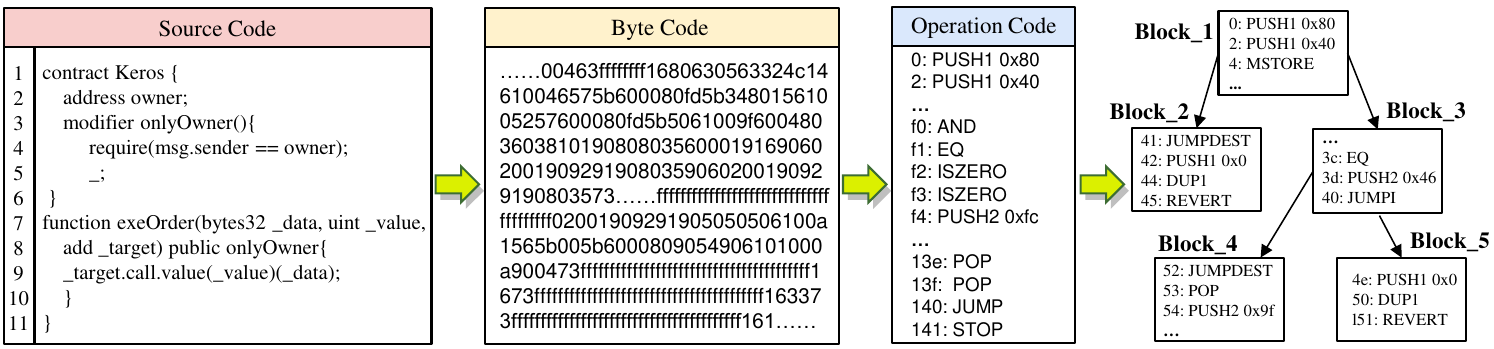}
    \caption{A simple illustrative example of constructing a Control Flow Graph (CFG) from Solidity source code. }
    \label{fig:source to cfg}
\end{figure*}

We segment the opcode stream into basic blocks, each comprising a linear sequence of instructions without internal jumps. A Control Flow Graph $G_f$ is then constructed, where each node $v_k \in V_f$ corresponds to a basic block, and each edge $e_{kl} \in E_f$ denotes a control-flow transition (e.g., conditional jumps, fall-through edges). Figure~\ref{fig:source to cfg} illustrates this process. The first pane shows a contract's source code, the second shows its compiled bytecode, the third lists its opcode sequence, and the fourth depicts the resulting CFG, consisting of blocks and control edges. To encode each block, we employ a pretrained representation model BERT\cite{vaswani2017attention} to generate contextual embeddings that capture both syntactic and semantic features of the opcodes. 

Compared to source code, bytecode offers more limited and lower-level semantic cues. Hence, constructing CFGs provides a means to abstract and organize bytecode logic in a structured format suitable for graph-based learning. By aligning this representation with that of CSGs, we facilitate subsequent cross-modal learning and knowledge transfer.
\subsection{Graph Encoding with Dual-Attention Graph Network (DAGN)}
\label{sec:graph_encoding}

To prevent vulnerability-critical nodes from being lost during graph aggregation, we design a \emph{Dual-Attention Graph Network} (DAGN) as the shared backbone for both modalities. DAGN introduces two mechanisms that act in sequence: a relation-aware GNN encoder that distinguishes edge types when aggregating neighborhood information, ensuring that control-flow edges carrying vulnerability-relevant semantics (e.g., external calls versus ordinary arithmetic) receive differentiated treatment; and an attention-based graph pooling layer that assigns learnable importance weights to individual nodes, allowing the few nodes that participate in vulnerability patterns to survive into the graph-level representation rather than being averaged away by the majority of normal nodes. Unless otherwise specified, the teacher (source-side) and student (bytecode-side) employ architecture-matched DAGN encoders.

\subsubsection{GNN Encoder}

Given a graph \(G=(V,E)\), DAGN updates node embeddings by attending over each node’s neighborhood with relation-aware attention. The node embedding update at the \( (l+1) \)-th layer is formalized as:
\begin{equation}
\label{eq:gnn_encoder}
\boldsymbol{\tilde{v}}_1^{(l+1)}, \dots, \boldsymbol{\tilde{v}}_{|V|}^{(l+1)} = f_\texttt{gnn}(\boldsymbol{v}_1^{(l)}, \dots, \boldsymbol{v}_{|V|}^{(l)}),
\end{equation}
where \( \boldsymbol{v}_i^{(l)} \) denotes the embedding of node \( i \) at layer \( l \). The update mechanism relies on aggregating information from each node’s local neighborhood using an attention mechanism:
\begin{equation}
\label{eq:node_update}
\boldsymbol{v}_i^{(l+1)} = f_v\left(\sum_{s \in \mathcal{N}_i \cup \{i\}} \alpha_{s,i} \boldsymbol{m}_{si}\right),
\end{equation}
where \( \mathcal{N}_i \) is the set of neighbors of node \( i \), \( \alpha_{s,i} \) denotes the attention coefficient for edge \( (s, i) \), and \( f_v(\cdot) \) represents a learnable feed-forward transformation. The message \( \boldsymbol{m}_{si} \) from node \( s \) to node \( i \) incorporates both the sender’s representation and the relation embedding:
\begin{equation}
\label{eq:message}
\boldsymbol{m}_{si} = f_m(\boldsymbol{v}_s^{(l)}, \boldsymbol{r}_{si}),
\end{equation}
where \( \boldsymbol{r}_{si} \) encodes the edge type between \( s \) and \( i \), and \( f_m(\cdot) \) is a learnable linear transformation. The attention weight \( \alpha_{s,i} \), which regulates the influence of incoming messages, is defined as:
\begin{equation}
\label{eq:attention}
\alpha_{s,i} = \text{Softmax}\left(\frac{\boldsymbol{q}_s \boldsymbol{k}_i^\top}{\sqrt{d}}\right),
\end{equation}
where the query vector \( \boldsymbol{q}_s = f_q(\boldsymbol{v}_s^{(l)}) \) and the key vector \( \boldsymbol{k}_i = f_k(\boldsymbol{v}_i^{(l)}, \boldsymbol{r}_{si}) \) are derived via learnable linear projections. Here, \( d \) denotes the dimensionality of the projection space.

\subsubsection{Attention-Based Graph Pooling (AGP)}
To derive a global graph representation suitable for downstream tasks, the node embeddings produced by the GNN encoder need to be aggregated into a fixed-size vector. Rather than relying on conventional aggregation methods \cite{qian2023cross_www,sun2025mtvhunter} that may suffer from oversmoothing or neglect semantically critical nodes, an attention-based pooling strategy is adopted:

Given the final-layer node embeddings \( \{\boldsymbol{v}_i^{(L)}\}_{i=1}^{|V|} \) from \( G = (V, E) \), the graph-level representation \( \boldsymbol{g} \in \mathbb{R}^d \) is computed through a weighted sum:
\begin{equation}
\label{eq:graph_pooling}
\boldsymbol{g} = AGP \left( \{\boldsymbol{v}_i^{(L)}\}_{i=1}^{|V|} \right) = \sum_{i=1}^{|V|} \beta_i \cdot \boldsymbol{v}_i^{(L)},
\end{equation}
where \( \beta_i \in [0, 1] \) is the attention weight indicating the importance of node \( i \) in capturing the overall semantics of the graph. These attention scores are defined as:
\begin{equation}
\label{eq:graph_attention}
\beta_i = \frac{\exp(\boldsymbol{w}^\top \tanh(W \boldsymbol{v}_i^{(L)} + b))}{\sum_{j=1}^{|V|} \exp(\boldsymbol{w}^\top \tanh(W \boldsymbol{v}_j^{(L)} + b))},
\end{equation}
where \( W \in \mathbb{R}^{d \times d} \), \( \boldsymbol{w} \in \mathbb{R}^{d} \), and \( b \in \mathbb{R}^d \) are learnable parameters. This pooling strategy enables the model to emphasize nodes with higher contextual relevance while suppressing less informative ones, thereby producing a more expressive graph-level embedding.

Through GNN Encoder and Attention-Based Graph Pooling, DAGN maps each modality-specific input graph to a compact vector that preserves informative local structures and supplies the structural basis for the downstream distillation objectives.

\subsection{Expert-Guided Vulnerability Pattern}
\label{sec:distillation}

\renewcommand\tabularxcolumn[1]{m{#1}}
\begin{table*}[t]
  \centering
  \caption{Expert Sub-patterns and Corresponding Key Nodes in Source and Bytecode Graphs, for Each Vulnerability.}
  \label{tab:Expert-Sub-patterns}
  \footnotesize
  \renewcommand{\arraystretch}{1.4}
  \renewcommand{\multirowsetup}{\centering\arraybackslash}
  \setlength{\tabcolsep}{4pt}
  \begin{tabularx}{\textwidth}{
      >{\centering\arraybackslash}m{1.8cm}|
      >{\centering\arraybackslash}m{3cm}|
      >{\raggedright\arraybackslash}X|
      >{\raggedright\arraybackslash}X
  }
    \noalign{\hrule height 1.2pt}
    \textbf{Vulnerability} & \textbf{Sub-pattern} & \multicolumn{1}{c|}{\textbf{Key Nodes in Source Graph}} & \multicolumn{1}{c}{\textbf{Key Nodes in Bytecode Graph}} \\
    \hline
    \multirow{3}{1.8cm}{Reentrancy}
      & callValueInvocation & Invocation of \texttt{call.value()}. & Block containing \texttt{CALLVALUE} instruction. \\
      \cdashline{2-4}
      & balanceDeduction & Variable updates immediately following \texttt{call.value()}. & Successor block of \texttt{CALLVALUE} contains an assignment operation. \\
      \cdashline{2-4}
      & enoughBalance & Balance sufficiency check before fund transfer. & Predecessor of \texttt{CALLVALUE} block contains a comparison operation. \\
    \hline
    \multirow{3}{1.8cm}{Timestamp Dependence} 
      & timestampInvocation & Use of \texttt{block.timestamp} or \texttt{block.number}. & Block containing \texttt{TIMESTAMP} or \texttt{BLOCKHASH} opcode. \\
      \cdashline{2-4}
      & timestampAssign & Assignment based on timestamp values. & Successor of timestamp block contains assignment to state or local variables. \\
      \cdashline{2-4}
      & timestampContamination & Propagation of timestamp value through data flow. & Successor of timestamp block contains value forwarding or arithmetic operations. \\
    \hline
    \multirow{3}{1.8cm}{Infinite Loop} 
      & loopStatement & Presence of \texttt{for}/\texttt{while} loop constructs. & Block contains \texttt{JUMP} with backward offset (destination $\leq$ source). \\
      \cdashline{2-4}
      & loopCondition & Loop exit condition check. & \texttt{PUSH 0x1} occurs before \texttt{JUMPI}, or absence of \texttt{SSTORE/SLOAD} in related blocks. \\
      \cdashline{2-4}
      & selfInvocation & Function invokes itself recursively or via delegate call. & Block contains \texttt{CALL}, \texttt{DELEGATECALL} or \texttt{STATICCALL} with no condition guarding the call. \\
    \noalign{\hrule height 1.2pt}
  \end{tabularx}
\end{table*}
\subsubsection{Dual-Modality Pattern Extraction}
Prior work \cite{liu2021interpretable_graph_expert_ame,sun2025mtvhunter} extracts expert-defined vulnerability patterns from source code, typically encoding them as one-hot vectors or heuristic scores. Yet source code is often unavailable at deployment, and prior research has not defined corresponding patterns for bytecode. Because source code and opcode (bytecode) differ substantially in structure and representation, we cannot assume the two share ready-made annotations; we must establish the correspondence explicitly. To this end, we summarize existing source-code patterns \cite{liu2021combining_expert_gnn,kalra2018zeus,luu2016making,jiang2018contractfuzzer} and design new bytecode counterparts for the three target vulnerabilities.

\paragraph{Reentrancy.} A reentrancy vulnerability arises when an external call (\texttt{call.value()}) allows the callee to re-enter the caller before its state update completes. We capture this through three sub-patterns---\emph{callValueInvocation}, \emph{balanceDeduction}, and \emph{enoughBalance}---which respectively mark the external call site, the subsequent balance update, and the balance-sufficiency precondition, in both source and bytecode. Their exact node-level criteria are listed in Table~\ref{tab:Expert-Sub-patterns}.

\paragraph{Timestamp Dependence.} This vulnerability stems from contracts that rely on block timestamps or block numbers to drive sensitive control flow. Our three sub-patterns---\emph{timestampInvocation}, \emph{timestampAssign}, and \emph{timestampContamination}---mark the timestamp access, its assignment to a variable, and its propagation into a condition, respectively, in both modalities. Node-level criteria are given in Table~\ref{tab:Expert-Sub-patterns}.

\paragraph{Infinite Loop.} An infinite loop lacks a proper termination condition, causing gas exhaustion. We define three sub-patterns---\emph{loopStatement}, \emph{loopCondition}, and \emph{selfInvocation}---that mark the loop construct, the loop exit condition, and any unconditional recursive invocation, in both source and bytecode. Node-level criteria are listed in Table~\ref{tab:Expert-Sub-patterns}.

\subsubsection{Annotation for CSG and CFG Node}

The encoding mechanisms above produce structured node embeddings, but they operate without knowing which nodes are vulnerability-relevant: prior cross-modal distillation aligns source and bytecode using only contract-level labels, leaving the model to discover vulnerability-corresponding regions through indirect gradient signals---a process that often fails when vulnerability patterns occupy only a small fraction of the graph. We make these correspondences explicit through the expert-guided annotations above. Based on the obtained expert patterns, we mark the key nodes in the two modality graphs of each contract, and every annotation in the source graph corresponds to one in the bytecode graph. According to the expert patterns defined in the previous section, the annotation methods of the key nodes of the three vulnerabilities are detailed in the table \ref{tab:Expert-Sub-patterns}.

For each contract \( C \), let \( G_s^C = (V_s^C, E_s^C) \) be the source graph (CSG), and \( G_f^C = (V_f^C, E_f^C) \) the bytecode graph (CFG). We define pattern-specific key node sets: \(
V_{s}^{p, C} \subseteq V_s^C, \quad V_{f}^{p, C} \subseteq V_f^C
\)
where \( p \in \mathcal{P} \) is a vulnerability pattern. 

To establish correspondence between the two graphs, we introduce a contract-specific alignment dictionary:
\begin{equation}
\label{eq:pattern_dict}
\begin{aligned}
\mathcal{D}^C = \left\{ i \mapsto j \;\middle|\; i \in \text{idx}(V_s^C),\, j \in \text{idx}(V_f^C),\, i \leftrightarrow j \text{ under } p \right\}
\end{aligned}
\end{equation}
This dictionary records the index-to-index mapping between semantically aligned nodes in \( G_s^C \) and \( G_f^C \) as determined by the pattern-specific annotations. Each contract therefore maintains its own dictionary \( \mathcal{D}^C \), enabling local correspondence.

The pattern dictionary \( \mathcal{D}^C \) and the corresponding annotated node sets \( V_s^{p, C} \) and \( V_f^{p, C} \) are used to augment training supervision. In addition to optimizing global graph-level objectives, we inject fine-grained guidance at the node/block level by enforcing pattern consistency and information flow along matched node pairs. The exact architecture and learning objectives leveraging these annotations are described in the next subsection.

\subsection{Cross-Modal Knowledge Distillation via Expert-Guided Dual-Focus Learning}
\label{sec:knowledge-distillation}

With vulnerability-critical nodes preserved by DAGN's attentive encoding and their cross-modal correspondences established by the pattern annotations, the remaining question is how to transfer knowledge from the frozen teacher (operating on source CSG) to the trainable student (operating on bytecode CFG). We draw on multimodal learning \cite{wang2023multi_with_miss} to propose a cross-modal knowledge distillation framework whose goal is to align the student's CFG representations with the teacher's semantically richer CSG representations in a shared latent space, so that the bytecode model recovers source-level semantics that would otherwise be inaccessible. This targets the third link of the chained failure directly: without a local objective, global alignment alone cannot enforce consistency where vulnerabilities manifest.

Our framework follows a teacher–student architecture, where the teacher operates on CSG and the student on CFG. Both models share the same GNN encoders and graph pooling networks introduced earlier, without additional distillation modules. This design reduces architectural complexity and improves scalability, while leveraging prior findings that direct distillation between GNNs can achieve effective knowledge transfer \cite{ji2021KD_prior,yang2023KD_prior}.

Let \( G_s^C \) and \( G_f^C \) denote the CSG and CFG of contract \( C \), respectively. The teacher model \( \mathcal{T} \) is first pretrained on \( G_s^C \) and kept frozen during distillation. The student model \( \mathcal{S} \), trained on \( G_f^C \), is supervised through a combination of global and local distillation objectives, both designed to align the semantic representations across modalities.

\paragraph{Global Semantic Distillation Loss.}

To retain whole-graph alignment, we define a global distillation loss that directly minimizes the discrepancy between the final graph-level embeddings produced by the teacher and student models. Let \( \boldsymbol{g}_s \in \mathbb{R}^d \) and \( \boldsymbol{g}_f \in \mathbb{R}^d \) be the output features after graph pooling from the teacher and student, respectively. The global loss is defined as:
\begin{equation}
\label{eq:global_distill_mse}
\begin{aligned}
\mathcal{L}_{{Global}}&=\left\| AGP \left( \{\boldsymbol{v}_i^{(L)}\}_{i=1}^{|V_{G_s}|} \right) - AGP \left( \{\boldsymbol{v}_i^{(L)}\}_{i=1}^{|V_{G_f}|} \right) \right\|_2^2 \\& = \left\| \boldsymbol{g}_s - \boldsymbol{g}_f \right\|_2^2
\end{aligned}
\end{equation}

\paragraph{Local Semantic Distillation Loss.}

Existing distillation aligns only whole-graph embeddings, leaving node-level correspondences between the two modalities unenforced. Leveraging the alignment dictionary \(\mathcal{D}^C\), we introduce a local semantic distillation loss that minimizes the Euclidean distance between expert-paired source CSG nodes and bytecode CFG nodes, enforcing region-level consistency at exactly the locations where vulnerabilities manifest. Given an annotated alignment set \( \mathcal{D}^C \) consisting of node pairs \( (i, j) \), where \( i \in V_s^C \) and \( j \in V_f^C \), the local loss is defined as:
\begin{equation}
\label{eq:local_distill}
\mathcal{L}_{{Local}} = \frac{1}{|\mathcal{D}^C|} \sum_{(i,j) \in \mathcal{D}^C} \left\| \boldsymbol{v}_{s_i}^{(L)} - \boldsymbol{v}_{f_j}^{(L)} \right\|_2^2
\end{equation}
The overall distillation objective jointly optimizes both global and local alignment without introducing additional balancing hyperparameters:
\begin{equation}
\label{eq:total_distill}
\mathcal{L}_{{Dist}}=Dist(\mathcal{T},\mathcal{S}) = \mathcal{L}_{{Global}} + \mathcal{L}_{{Local}}
\end{equation}
After distillation, the teacher model is discarded. The student model is then fine-tuned under supervised learning to perform binary classification of contract vulnerability. The final prediction is obtained by applying a multi-layer perceptron (MLP) followed by softmax normalization to the graph-level representation \( \boldsymbol{g}_f \) produced by the student model. The supervised learning objective is defined as:
\begin{equation}
\label{eq:supervised_loss_softmax}
\mathcal{L}_{\text{Sup}} = -\sum_{c=1}^{2} y_c \log \left( \text{Softmax} \left( \text{MLP}(\boldsymbol{g}_f) \right)_c \right)
\end{equation}
Here, \( \boldsymbol{y}\) is the one-hot encoded ground-truth label vector, and the index \( c \) iterates over the vulnerability and normal classes.

\section{Evaluation}

We conduct a comprehensive evaluation to validate the effectiveness of our proposed framework. Specifically, we seek to address the following research questions:

\begin{itemize}
    \item \textbf{RQ1}: Can our proposed approach effectively detect three common types of smart contract vulnerabilities? How does it perform compared to traditional static analysis methods?
    \item \textbf{RQ2}: How does our method perform in comparison to existing deep learning-based approaches?
    \item \textbf{RQ3}: Does the proposed teacher-student knowledge distillation strategy enhance the model’s performance?
    \item \textbf{RQ4}: What is the contribution of the proposed graph pooling moudle to the overall framework?
    \item \textbf{RQ5}:  Do the expert vulnerability patterns and expert-guided alignment distillation strategies we designed effectively enhance the model?
     
\end{itemize}

We first introduce the dataset and experimental settings before presenting detailed results and analysis corresponding to each research question.

\subsection{Dataset and Experimental Setup}
\paragraph{Dataset}
We obtained public smart contracts based on the methods published by Qian et al. \cite{qian2023cross_www,liu2021interpretable_graph_expert_ame}. The dataset covers a variety of vulnerability types, including reentrancy vulnerabilities, timestamp dependency vulnerabilities, and infinite loop vulnerabilities. These contracts were collected from multiple sources such as the Ethereum platform, GitHub repositories, and technical blog posts analyzing real-world contracts. Each contract includes its Solidity source code and compiled bytecode.

To ensure data quality and consistency, several preprocessing steps were applied: (1) contracts with missing source code or bytecode were removed, (2) simple and trivial contracts were filtered out, and (3) contracts with unmatched or corrupted source–bytecode mappings were excluded. After cleaning, we retained 273 samples labeled with reentrancy vulnerabilities, 349 with timestamp dependency vulnerabilities, and 196 with infinite loop vulnerabilities. Each sample was annotated with a binary ground-truth label indicating the presence or absence of a vulnerability.  To support robust training and evaluation, we adopt stratified random sampling to partition the dataset into training, validation, and test sets with a fixed ratio of 7:1:2. All experiments were conducted over five independent runs with different random splits, and the average results are reported to ensure statistical reliability.

\paragraph{Experimental Setup}

All experiments were conducted on a server equipped with NVIDIA RTX 3090 GPUs (24GB memory), running Ubuntu 20.04.01 LTS with NVIDIA Driver Version 535.171.0. The implementation was developed in Python 3.8.10 using PyTorch 2.3.1 and CUDA 12.2. For optimization, we employed the Adam optimizer throughout all training procedures. The learning rate was selected via grid search from the set \(\{1e\text{-}4, 5e\text{-}4, 1e\text{-}3, 5e\text{-}3\}\). Our graph neural network architecture uses two layers for both the contract structure graph (CSG) and control flow graph (CFG) encoders. All hidden representations are 128-dimensional. A mini-batch size of 64 is used during training. We evaluate the classification performance using four standard metrics in binary classification: accuracy, precision, recall, and F1-score. 


\begin{table*}[ht]
    \centering
    \small
    \caption{Performance comparison (\%) between our model and traditional tools, ‘n/a’ means the corresponding tool does not support detecting the vulnerability type.}
    \resizebox{1\linewidth}{!}{
    \renewcommand{\arraystretch}{1}
    \begin{tabular}{c|cccccccccccc}
    \toprule[1.1pt]
        & \multicolumn{4}{c}{Reentrancy} & \multicolumn{4}{c}{Timestamp Dependence} & \multicolumn{4}{c}{Infinite Loop}  \\ \cmidrule{2-13}
        Methods     & Acc & Prec & Rec & F1 & Acc & Prec & Rec & F1 & Acc & Prec & Rec & F1 \\
    \cmidrule(lr){1-1} \cmidrule(lr){2-5} \cmidrule(lr){6-9} \cmidrule(lr){10-13}
    Securify \cite{tsankov2018securify}& 71.89 & 50.85 & 56.60 & 53.57 & n/a   & n/a   & n/a   & n/a   & n/a & n/a & n/a & n/a \\
    Smartchecker \cite{tikhomirov2018smartcheck}& 52.97 & 25.00 & 32.08 & 28.10 & 44.32 & 39.16 & 37.25 & 38.18 & n/a & n/a & n/a & n/a \\
    Oyente \cite{luu2016making}& 61.62 & 38.16 & 54.71 & 44.96 & 59.45 & 45.16 & 38.44 & 41.53 & n/a & n/a & n/a & n/a \\
    Mythril \cite{mueller2017framework}& 60.54 & 39.58 & 71.69 & 51.02 & 61.08 & 50.00 & 41.72 & 45.49 & n/a & n/a & n/a & n/a \\
    Slither \cite{feist2019slither}& 77.12 & 68.42 & 74.28 & 71.23 & 74.20 & 67.25 & 72.38 & 69.72 & n/a & n/a & n/a & n/a \\
    PDA  \cite{ibing2015pda} & n/a   & n/a   & n/a   & n/a   & n/a   & n/a   & n/a   & n/a   & 46.44 & 42.96 & 21.73 & 28.26 \\
    Looper \cite{burnim2009looper}& n/a   & n/a   & n/a   & n/a   & n/a   & n/a   & n/a   & n/a   & 59.56 & 62.72 & 47.21 & 53.87 \\
    \midrule
    \cellcolor{Gray}Ours & \cellcolor{Gray}\textbf{90.72} & \cellcolor{Gray}\textbf{92.22} & \cellcolor{Gray}\textbf{88.58} & \cellcolor{Gray}\textbf{90.78} & \cellcolor{Gray}\textbf{89.87} & \cellcolor{Gray}\textbf{89.08} & \cellcolor{Gray}\textbf{91.40} & \cellcolor{Gray}\textbf{90.23} & \cellcolor{Gray}\textbf{83.11} & \cellcolor{Gray}\textbf{82.74} & \cellcolor{Gray}\textbf{85.06} & \cellcolor{Gray}\textbf{83.94} \\
    \bottomrule[1.1pt]
    \end{tabular}}
\label{tab:compare with tradition}
\end{table*}

\begin{table*}[ht]
    \centering
    \small
    \caption{Performance comparison (\%) between our model and learning based methods, The 'Improve' row reports the relative improvement (\%) of our model over the best-performing baseline.}
    \resizebox{1\linewidth}{!}{
    \renewcommand{\arraystretch}{1}
    \begin{tabular}{c|cccccccccccc}
    \toprule[1.1pt]
        & \multicolumn{4}{c}{Reentrancy} & \multicolumn{4}{c}{Timestamp Dependence} & \multicolumn{4}{c}{Infinite Loop}  \\ \cmidrule{2-13}
        Methods     & Acc & Prec & Rec & F1 & Acc & Prec & Rec & F1 & Acc & Prec & Rec & F1 \\
    \cmidrule(lr){1-1} \cmidrule(lr){2-5} \cmidrule(lr){6-9} \cmidrule(lr){10-13}
    Vanilla-RNN \cite{tann2018vanilla-rnn}& 49.64 & 49.82 & 58.78 & 50.71 & 49.77 & 51.91 & 44.59 & 45.62 & 49.57 & 42.10 & 47.86 & 44.79 \\
    LSTM \cite{sak2014lstm} & 53.68 & 51.65 & 67.82 & 58.64 & 50.79 & 50.32 & 59.23 & 54.41 & 51.28 & 44.07 & 57.26 & 49.80 \\
    GRU  \cite{chung2014gru} & 54.54 & 53.10 & 71.30 & 60.87 & 52.06 & 49.41 & 59.91 & 54.15 & 51.70 & 45.00 & 50.42 & 47.55 \\
    GCN \cite{kipf2016gcn}  & 77.85 & 70.02 & 78.79 & 74.15 & 74.21 & 68.35 & 75.97 & 71.96 & 64.01 & 59.96 & 63.04 & 61.46 \\
    TMP  \cite{zhuang2021smart_tmp} & 84.48 & 74.06 & 82.63 & 78.11 & 83.45 & 75.05 & 83.82 & 79.19 & 74.61 & 73.89 & 74.32 & 74.10 \\
    AME \cite{liu2021interpretable_graph_expert_ame}  & 90.19 & 86.25 & 89.69 & 87.94 & 86.52 & 82.07 & 86.23 & 84.10 & 80.32 & 78.69 & 79.08 & 78.88 \\
    SMS \cite{qian2023cross_www}  & 88.94 & 86.00 & 89.53 & 88.14 & 87.28 & 84.29 & 88.56 & 86.35 & 79.15 & 77.93 & 77.54 & 77.72 \\
    \midrule
    \cellcolor{Gray}Ours & \cellcolor{Gray}\textbf{90.72} & \cellcolor{Gray}\textbf{92.22} & \cellcolor{Gray}\textbf{88.58} & \cellcolor{Gray}\textbf{90.86} & \cellcolor{Gray}\textbf{89.87} & \cellcolor{Gray}\textbf{89.08} & \cellcolor{Gray}\textbf{91.40} & \cellcolor{Gray}\textbf{90.23} & \cellcolor{Gray}\textbf{83.11} & \cellcolor{Gray}\textbf{82.74} & \cellcolor{Gray}\textbf{85.06} & \cellcolor{Gray}\textbf{83.94} \\
    \cellcolor{Gray}\textcolor{bgreen}{Improv.} (\%) & \cellcolor{Gray}0.59  & \cellcolor{Gray}6.92  & \cellcolor{Gray}-1.24  & \cellcolor{Gray}3.09  & \cellcolor{Gray}2.97  & \cellcolor{Gray}5.68  & \cellcolor{Gray}3.21  & \cellcolor{Gray}4.49  & \cellcolor{Gray}3.47  & \cellcolor{Gray}5.15  & \cellcolor{Gray}7.56  & \cellcolor{Gray}6.41  \\
    \bottomrule[1.1pt]
    \end{tabular}}
\label{tab:compare with learing base}
\end{table*}

\subsection{Comparison with Traditional Tools (RQ1)}

To evaluate the effectiveness of our proposed method in vulnerability detection, we conduct a comprehensive comparison with several representative baseline tools. Specifically, we select the following state-of-the-art traditional static and symbolic analysis tools: \textbf{Securify} \cite{tsankov2018securify}, a formal verification-based tool; \textbf{SmartChecker} \cite{tikhomirov2018smartcheck}, which leverages abstract syntax trees for static analysis; \textbf{Oyente} \cite{luu2016making}, one of the earliest and widely recognized symbolic execution engines; \textbf{Mythril} \cite{mueller2017framework}, which combines concolic analysis, taint tracking, and control flow checking; \textbf{Slither} \cite{feist2019slither}, which transforms smart contracts into an intermediate representation to identify vulnerabilities; \textbf{PDA} \cite{ibing2015pda}, which performs program path checking; and \textbf{Looper} \cite{burnim2009looper}, a symbolic execution-based detector for control-flow related vulnerabilities. These tools have been widely adopted in previous studies, offering a strong and diverse baseline for evaluating the performance of our approach.
The experimental results are summarized in Table~\ref{tab:compare with tradition}. 

For reentrancy detection, our method achieves an F1-score of 90.78, significantly outperforming the best baseline, Slither (71.23). Although Slither benefits from IR-based representations and heuristic rules, it is limited by rule coverage and generalization. In contrast, our model captures complex interprocedural patterns, improving all metrics with a +19.55 gain in F1-score. For timestamp dependence, our approach again outperforms all baselines, achieving an F1-score of 90.23 compared to Slither’s 69.72. This improvement indicates a stronger ability to capture temporal logic flaws, especially when timestamps are indirectly involved in control or state updates. For infinite loop detection, which requires control-flow-intensive analysis, our model achieves an F1-score of 83.94, surpassing Looper (53.87), the strongest baseline for this category.

Overall, our method consistently outperforms existing tools across all metrics, with improvements of up to 25.57\% in precision, 8.28\% in recall, and 12.86\% in accuracy over the best baselines.  We attribute the performance gap primarily to the inherent limitations of traditional static and symbolic analysis tools. For example, Mythril detects reentrancy by matching specific call patterns (e.g., \texttt{call} without proper state updates), but may miss indirect or multi-level cases. In contrast, our data-driven approach captures such behaviors implicitly, improving recall and reducing false negatives, thereby demonstrating strong generalization across vulnerability types.

\subsection{Compare with Learning Based Methods (RQ2)}

To evaluate the effectiveness of our method from a learning-based perspective, we compare it with a series of representative neural approaches for smart contract vulnerability detection. These include both sequence modeling and graph-based methods, covering a wide range of perspectives in learning-based vulnerability detection.

We consider \textbf{Vanilla-RNN} \cite{tann2018vanilla-rnn}, which takes contract function code sequences as input and captures sequential patterns through hidden state propagation. \textbf{LSTM} \cite{sak2014lstm} and \textbf{GRU} \cite{chung2014gru} extend this approach with gating mechanisms to improve the modeling of long-range dependencies. For graph-based methods, we include \textbf{GCN} \cite{kipf2016gcn}, which applies graph convolution operations over structured contract representations; \textbf{TMP} \cite{zhuang2021smart_tmp}, which adopts a temporal message passing mechanism to model information flow; and \textbf{AME} \cite{liu2021interpretable_graph_expert_ame}, which integrates expert patterns into attention-guided neural models. Additionally, we include \textbf{SMS} \cite{qian2023cross_www}, a method designed to detect vulnerabilities directly from bytecode by learning source code semantic patterns through mutual learning. In particular, in order to make feasible comparisons, Vanilla-RNN, LSTM, and GRU are provided with contract function source code sequences, TMP, AME extracts normalized graphs from the source code and follows their published implementations.

The evaluation results are shown in Table~\ref{tab:compare with learing base}. Our method achieves the highest F1-scores across all three types of vulnerabilities—reentrancy, timestamp dependence, and infinite loop. Notably, despite relying solely on bytecode during the detection phase, our method achieves an F1-score of 90.86 for reentrancy, 90.23 for timestamp dependence, and 83.94 for infinite loops. Compared to the strongest baseline, these results correspond to absolute F1 improvements of 2.72\%, 3.88\%, and 6.22\%, respectively. As shown in the last row of Table~\ref{tab:compare with learing base}, these improvements also represent relative F1-score gains of 3.09\%, 4.49\%, and 6.41\% across the three vulnerability types.

Among the sequence-based models, Vanilla-RNN, LSTM, and GRU exhibit noticeably lower performance across all vulnerability types. In many cases, they underperform even traditional non-learning-based tools. This highlights a fundamental limitation in treating smart contract code as linear token sequences: such models struggle to capture the complex control and data dependencies essential for accurate vulnerability detection. For example, their F1-scores on Infinite Loop remain low (Vanilla-RNN 44.79, LSTM 49.80, GRU 47.55), and even on Reentrancy they lag behind (Vanilla-RNN 50.71, LSTM 58.64, GRU 60.87). In contrast, graph-based models like GCN and TMP yield considerably stronger results, demonstrating the utility of structural representations (e.g., on Reentrancy, GCN 74.15 and TMP 78.11). However, their reliance on normalized source-level graphs and the absence of explicit domain supervision may limit their practicality, particularly under scenarios involving incomplete or obfuscated contracts. SMS, by contrast, performs well under these constraints, benefiting from its knowledge transfer from source code to bytecode design (e.g., F1 on Reentrancy 88.14), though its reliance on coarse pooling and the lack of explicit expert guidance leave room for improvement. We speculate that insufficiently expressive aggregation mechanisms, the absence of domain-informed supervisory signals, and the lack of objectives that balance global representation learning with fine-grained discrimination may all contribute to the performance gaps observed in prior approaches.

\begin{table*}[t]
    \centering
    \small
    \caption{Performance comparison (\%) among different distillation variants and our full model. The 'Dec.' rows indicate the relative performance decrease (\%) of D-GNN and D-AGP compared to the full model for each metric. The 'Improv.' row reports the relative improvement (\%) of our full model over the variant without distillation.}
    \resizebox{0.95\linewidth}{!}{
    \renewcommand{\arraystretch}{1}
    \begin{tabular}{c|cccccccccccc}
    \toprule[1.1pt]
        & \multicolumn{4}{c}{Reentrancy} & \multicolumn{4}{c}{Timestamp Dependence} & \multicolumn{4}{c}{Infinite Loop}  \\ \cmidrule{2-13}
        Methods     & Acc & Prec & Rec & F1 & Acc & Prec & Rec & F1 & Acc & Prec & Rec & F1 \\
    \cmidrule(lr){1-1} \cmidrule(lr){2-5} \cmidrule(lr){6-9} \cmidrule(lr){10-13}
    D-GNN & 89.09 & 90.28 & 84.59 & 87.79 & 89.39 & 88.16 & 88.25 & 87.7  & 82.05 & 83.15 & 80.88 & 80.72 \\
    \textcolor{bred}{Dec.} (\%) & 1.80  & 2.10  & 4.50  & 3.29  & 0.53  & 1.03  & 3.45  & 2.80  & 1.28  & -0.50  & 4.91  & 3.84  \\
    \hdashline
    D-AGP & 90.09 & 91.02 & 80.7  & 86.06 & 88.2  & 85.87 & 89.52 & 86.44 & 78.95 & 78.12 & 80.33 & 78.69 \\
   \textcolor{bred}{ Dec.} (\%) & 0.69  & 1.30  & 8.90  & 5.20  & 1.86  & 3.60  & 2.06  & 4.20  & 5.01  & 5.58  & 5.56  & 6.25  \\
    \midrule
    w/o Distill & 87.82 & 86.69 & 81.05 & 84.06 & 86.6  & 83.75 & 86.5  & 84.73 & 80.33 & 79.21 & 82.85 & 79.93 \\
    \hdashline
    \cellcolor{Gray}Ours & \cellcolor{Gray}\textbf{90.72} & \cellcolor{Gray}\textbf{92.22} & \cellcolor{Gray}\textbf{88.58} & \cellcolor{Gray}\textbf{90.78} & \cellcolor{Gray}\textbf{89.87} & \cellcolor{Gray}\textbf{89.08} & \cellcolor{Gray}\textbf{91.40} & \cellcolor{Gray}\textbf{90.23} & \cellcolor{Gray}\textbf{83.11} & \cellcolor{Gray}\textbf{82.74} & \cellcolor{Gray}\textbf{85.06} & \cellcolor{Gray}\textbf{83.94}\\
    \cellcolor{Gray}\textcolor{bgreen}{Improv.} (\%) & \cellcolor{Gray}3.30  & \cellcolor{Gray}6.38  & \cellcolor{Gray}9.29  & \cellcolor{Gray}7.99  & \cellcolor{Gray}3.78  & \cellcolor{Gray}6.36  & \cellcolor{Gray}5.66  & \cellcolor{Gray}6.49  & \cellcolor{Gray}3.46  & \cellcolor{Gray}4.46  & \cellcolor{Gray}2.67  & \cellcolor{Gray}5.02  \\
    \bottomrule[1.1pt]
    \end{tabular}}
\label{tab:eval of distill}
\end{table*}

\begin{table*}[ht]
    \centering
    \small
    \caption{Performance comparison (\%) of different graph pooling strategies, including average pooling, max pooling, power pooling, and our proposed AGP module. The 'Dec.' rows indicate the relative performance decrease (\%) of each alternative pooling method compared to our full model with AGP for each metric.}
    \resizebox{0.95\linewidth}{!}{
    \renewcommand{\arraystretch}{1}
    \begin{tabular}{c|cccccccccccc}
    \toprule[1.1pt]
        & \multicolumn{4}{c}{Reentrancy} & \multicolumn{4}{c}{Timestamp dependence} & \multicolumn{4}{c}{Infinite Loop}  \\ \cmidrule{2-13}
        Methods     & Acc & Prec & Rec & F1 & Acc & Prec & Rec & F1 & Acc & Prec & Rec & F1 \\
    \cmidrule(lr){1-1} \cmidrule(lr){2-5} \cmidrule(lr){6-9} \cmidrule(lr){10-13}
    Avg & 89.32 & 90.25 & 85.00 & 86.77 & 86.74 & 84.00 & 87.28 & 85.58 & 80.62 & 77.43 & 81.79 & 79.56 \\
    \textcolor{bred}{Dec.} (\%) & 1.54  & 2.14  & 4.04  & 4.42  & 3.48  & 5.70  & 4.51  & 5.15  & 3.00  & 6.42  & 3.84  & 5.22  \\
    \hdashline
    Max & 88.13 & 89.67 & 82.35 & 84.59 & 84.03 & 86.40 & 81.80 & 84.03 & 79.48 & 80.88 & 75.85 & 78.27 \\
    \textcolor{bred}{Dec.} (\%) & 2.85  & 2.77  & 7.03  & 6.82  & 6.50  & 3.01  & 10.50  & 6.87  & 4.37  & 2.25  & 10.83  & 6.75  \\
    \hdashline
    Power & 89.85 & 91.83 & 85.44 & 87.31 & 88.31 & 88.36 & 86.70 & 87.54 & 81.47 & 80.39 & 81.87 & 81.13 \\
    \textcolor{bred}{Dec.} (\%) & 0.96  & 0.42  & 3.54  & 3.82  & 1.74  & 0.81  & 5.14  & 2.98  & 1.97  & 2.84  & 3.75  & 3.35  \\
    \midrule
    \cellcolor{Gray}Ours & \cellcolor{Gray}\textbf{90.72} & \cellcolor{Gray}\textbf{92.22} & \cellcolor{Gray}\textbf{88.58} & \cellcolor{Gray}\textbf{90.78} & \cellcolor{Gray}\textbf{89.87} & \cellcolor{Gray}\textbf{89.08} & \cellcolor{Gray}\textbf{91.40} & \cellcolor{Gray}\textbf{90.23} & \cellcolor{Gray}\textbf{83.11} & \cellcolor{Gray}\textbf{82.74} & \cellcolor{Gray}\textbf{85.06} & \cellcolor{Gray}\textbf{83.94} \\
    \bottomrule[1.1pt]
    \end{tabular}}
\label{tab:deanrole_baseline_with_errors}
\end{table*}

\subsection{Evaluation of Distillation Learning (RQ3)}
One of the main objectives of our work is to facilitate cross-modal information fusion and transfer between source code and bytecode representations of smart contracts via knowledge distillation. Rather than adding separate recovery networks, our approach aligns high-level vulnerability patterns directly with both the graph encoder and the attention-based graph pooling component. To investigate whether our teacher-student knowledge distillation strategy enhances the model's performance, we introduce three controlled variants of our model. In the first variant (D-GNN), the distillation loss only supervises the graph neural network encoder while freezing the parameters of the attention-based pooling module. The second variant (D-AGP) applies the distillation loss solely to the pooling module, keeping the GNN encoder fixed. The third variant (w/o Distill) disables the distillation process entirely, and excluding expert patterns from supervision. All other training conditions remain unchanged to ensure fair comparison.

As shown in Table~\ref{tab:eval of distill}, both D-GNN and D-AGP experience performance drops compared to the full model. For instance, on reentrancy detection, D-GNN exhibits a relative F1-score decline of 3.29\%, while D-AGP suffers a larger reduction of 5.20\%. This trend is consistent across other vulnerability types, where D-AGP shows greater degradation in most metrics. One plausible explanation for this outcome is that freezing the GNN encoder in D-AGP prevents the local distillation loss from effectively optimizing the intermediate node-level representations. As a result, the trasfer of semantic knowledge into the overall contract encoding becomes less effective, which negatively impacts downstream detection.

Furthermore, when compared to the variant without distillation, our full model exhibits consistent improvements across all evaluated metrics and vulnerability types. Specifically, we observe absolute F1-score gains of 7.99\% for reentrancy, 6.49\% for timestamp dependence, and 5.02\% for infinite loop vulnerabilities. These results suggest that distilling information from source code provides meaningful benefits in enhancing the bytecode-based model’s ability to detect vulnerabilities. Since source code contains high-level semantic constructs—such as abstract control structures and descriptive identifiers—that are often lost during compilation, the teacher model can serve as a valuable auxiliary signal for the student. By aligning internal representations across modalities, the distillation process help compensate for semantic loss in the bytecode, enabling more robust and generalizable detection. 

\subsection{Evaluation on DAGN moudle (RQ4)}

To assess the contribution of our proposed Attention-based Graph Pooling (AGP) module in DAGN network to the overall framework, we conduct an ablation study by replacing AGP with three alternative pooling strategies: average pooling, max pooling, and power pooling. These variants are designed to evaluate how different aggregation mechanisms over contract graph node features affect the final performance of our model.

In the Avg Pooling variant, we replace AGP with standard average pooling, where node features in the contract graph are aggregated by computing their mean to obtain a graph-level representation. In the Max Pooling variant, the model selects the maximum value across each node feature dimension. Power Pooling combines pooling with an element-wise power transformation, modulated by a learnable exponent parameter $p$, initialized to 1.0. This mechanism allows for a tunable balance between emphasizing dominant node features and preserving the overall distribution.

Table~\ref{tab:deanrole_baseline_with_errors} reports the performance of these model variants on the reentrancy, timestamp dependence, and infinite loop detection tasks. For clarity, we report both absolute values and relative degradation in key metrics compared to the full model using AGP. Across all tasks and metrics, replacing AGP with alternative pooling schemes leads to a measurable decline in performance. The degradation is most pronounced with max pooling. For instance, on the reentrancy task, the F1-score drops by 6.82\% relative to the AGP-based model. A similar trend is observed in recall across all tasks, indicating that max pooling may fail to retain sufficient global semantic context from the contract graph. The power pooling variant shows the least performance decline overall. This may be attributed to the presence of the learnable exponent, which allows the model to dynamically adjust the emphasis on feature magnitude during aggregation. In contrast, both average and max pooling apply fixed aggregation rules, which may underutilize the rich semantics encoded in graph node features.

\subsection{Evaluation of designed expert pattern(RQ5)}
\begin{figure}[ht]
    \centering
    \begin{minipage}[b]{0.24\textwidth}
        \centering        \includegraphics[width=\textwidth]{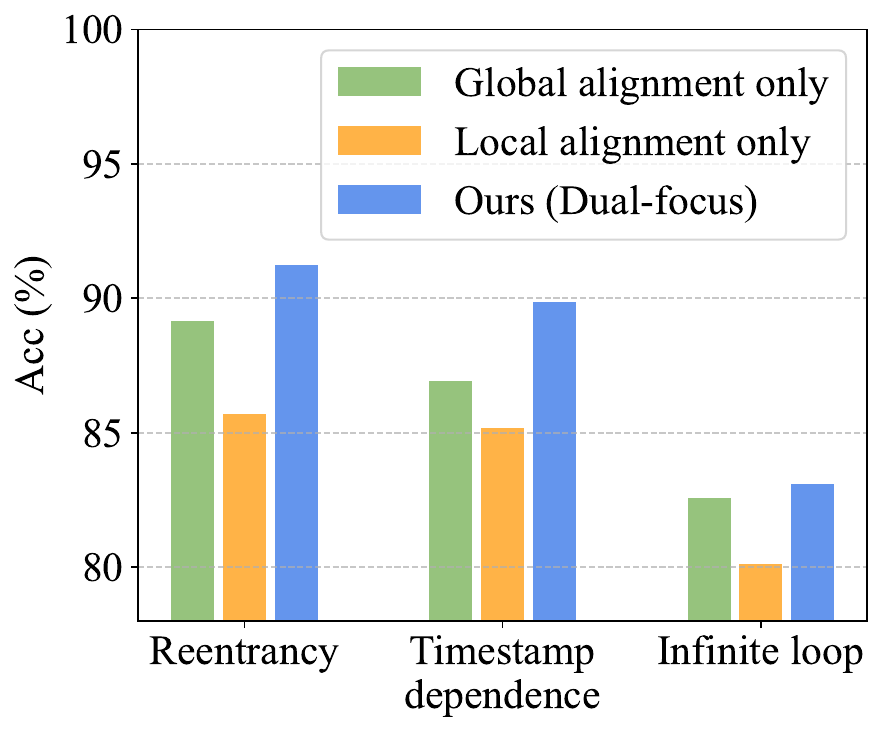}
    \end{minipage}
    \begin{minipage}[b]{0.24\textwidth}
        \centering
        \includegraphics[width=\textwidth]{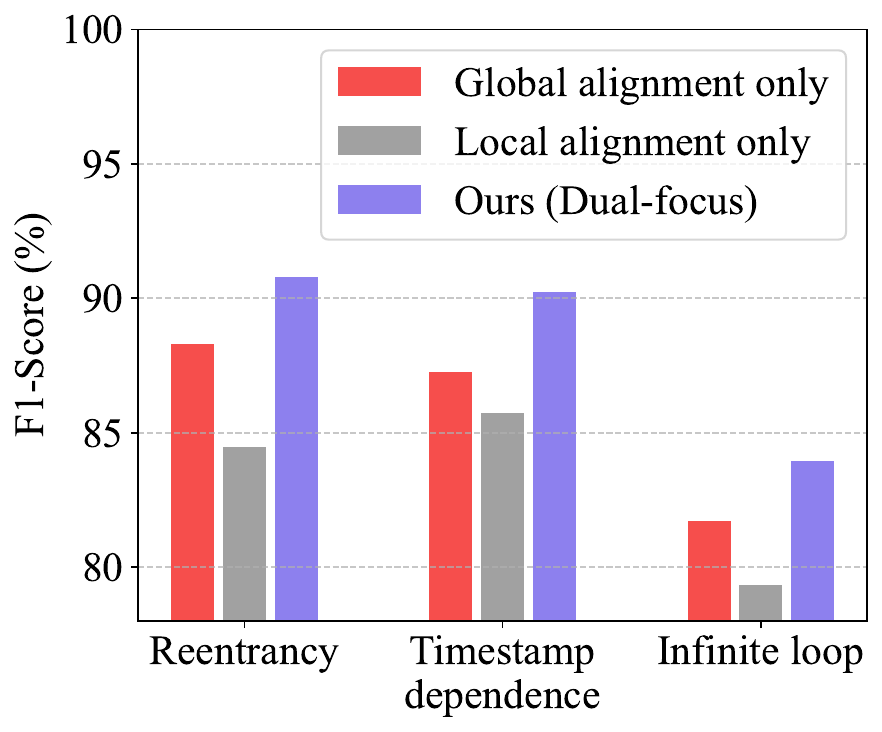}
    \end{minipage}
    \caption{Performance comparison of the student models under different alignment losses during distillation. The left chart reports accuracy, and the right reports F1-score.}
    \label{fig:expert_ablation_main}
\end{figure}

To examine the contribution of the expert pattern module to our overall framework, particularly its role in guiding the dual-focus distillation strategy during training, we perform an ablation study that targets the loss components involved in knowledge transfer. Specifically, during the distillation phase, our method introduces two complementary objectives: a global-level semantic alignment loss at the graph level and a local-level semantic alignment loss at the node level. To isolate the influence of each component, we construct two model variants, each removing one of the two losses while keeping the other intact. After distillation, the resulting student models are further trained under the same supervised fine-tuning protocol as the full model, and their performance is evaluated accordingly.

Figure~\ref{fig:expert_ablation_main} shows the accuracy and F1-score comparisons across three vulnerability types: reentrancy, timestamp dependence, and infinite loop. The bars labeled “Global alignment only” correspond to models trained with only the graph-level loss, while “Local alignment only” refers to models trained with only the node-level loss. “Ours (Dual-focus)” indicates the full model trained with both alignment signals. The results demonstrate that removing either loss component results in a noticeable degradation in performance across all tasks. For instance, with reentrancy, F1 under local-only is  84.5\%, under global-only is  88.3\%, whereas the dual-focus model reaches about 90.78\%. A similar trend holds for timestamp dependence (local-only 85\%, global-only 87\%, dual-focus 91\%) and for infinite loop (local-only 79\%, global-only 82\%, dual-focus 84\%). These observations suggest that local alignment alone may not capture sufficient global context or contract-level semantics required for robust vulnerability identification. Conversely, the model trained with only the global alignment loss performs better than its local-only counterpart, but still falls short of the full model, indicating that node-level cues also provide essential fine-grained supervision. These results underscore the necessity of enforcing semantic consistency across both holistic and structural dimensions to achieve robust generalization.

\begin{figure*}[t]
    \centering
    \begin{minipage}[b]{0.32\textwidth}
        \centering
        \includegraphics[width=\textwidth]{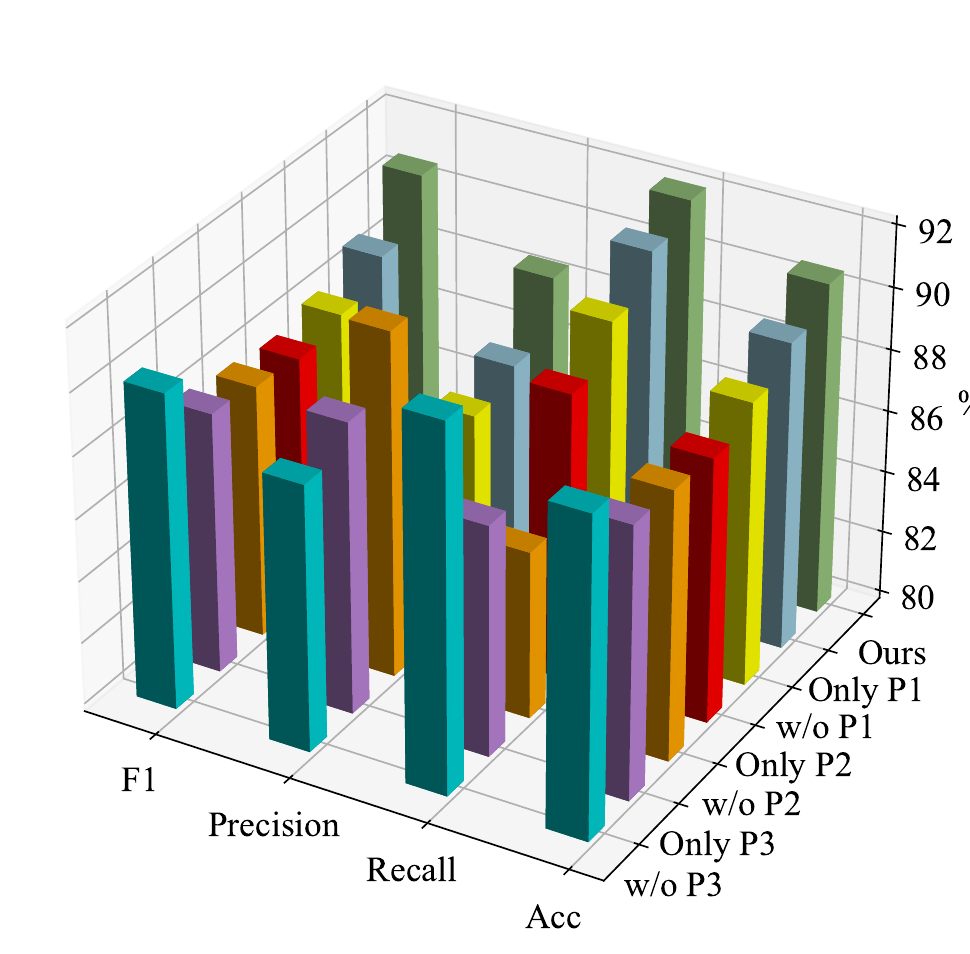}
        \subcaption{\footnotesize Reentrancy}
    \end{minipage}
    \begin{minipage}[b]{0.32\textwidth}
        \centering
        \includegraphics[width=\textwidth]{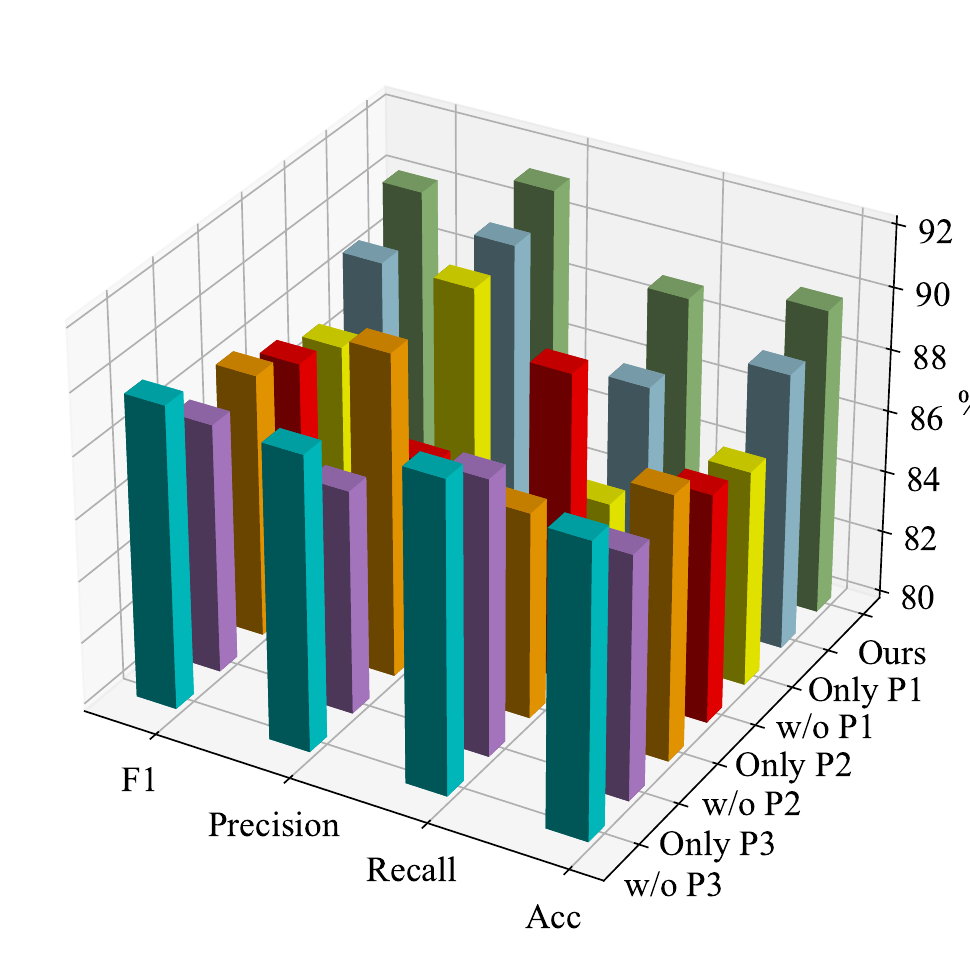}
        \subcaption{\footnotesize Timestamp dependence}
    \end{minipage}
    \begin{minipage}[b]{0.32\textwidth}
        \centering
        \includegraphics[width=\textwidth]{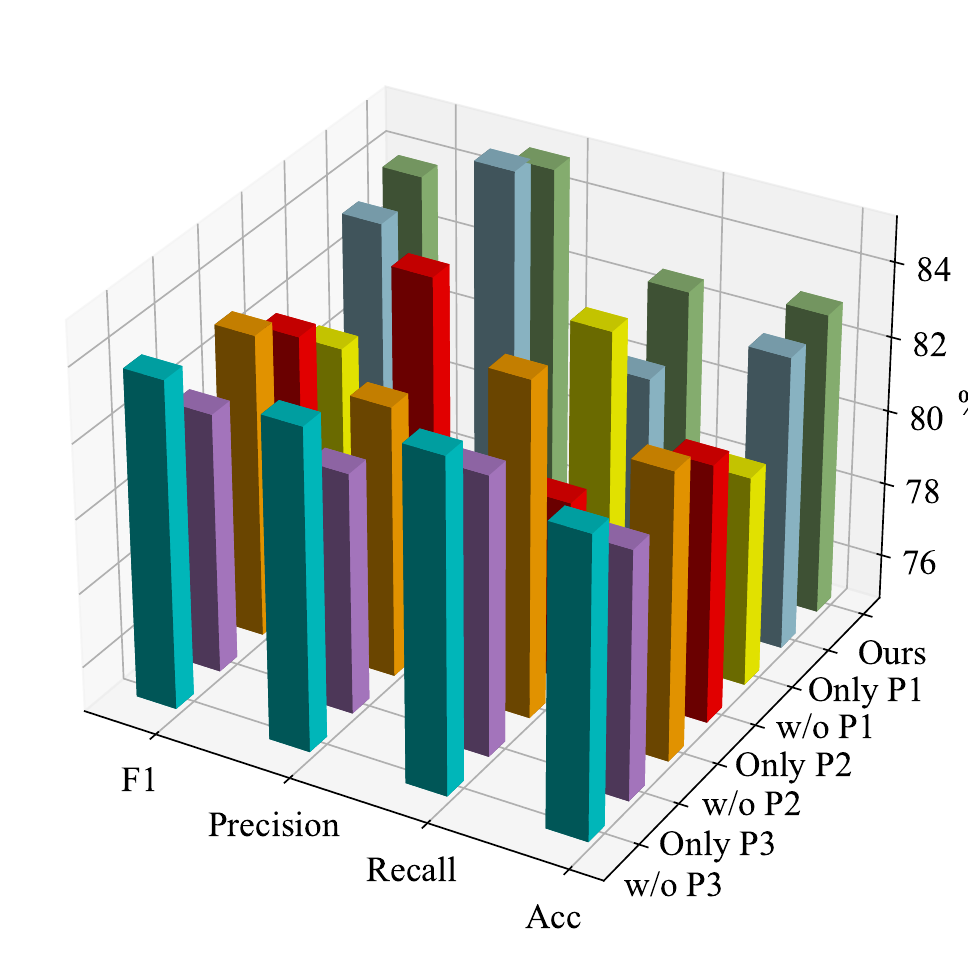}
        \subcaption{\footnotesize Infinite loop}
    \end{minipage}
    \caption{Performance under various expert sub-pattern ablation settings. Each 3D bar chart reports Accuracy, Recall, Precision, and F1-score when selectively including or excluding specific sub-patterns for each vulnerability type.}
    \label{fig:expert_ablation_subpattern}
\end{figure*}

To further investigate the role of structured expert knowledge in guiding local alignment, we conduct an additional ablation study focused on the sub-patterns designed for each vulnerability type. As outlined in the methodology section, we define three expert sub-patterns respectively for reentrancy, timestamp dependence, and infinite loop vulnerabilities. These sub-patterns are designed to annotate specific nodes (code blocks) in the contract graph, thereby enabling localized alignment between expert knowledge and model representation. To assess the individual contribution of each sub-pattern, we design two types of ablation studies per vulnerability: (i) removing one specific sub-pattern while retaining the others, and (ii) preserving only one sub-pattern and removing the rest. For instance, in the case of reentrancy, the configuration w/o P2 denotes the exclusion of the second sub-pattern (balanceDeduction), and no nodes will be marked with this pattern. Conversely, Only P3 refers to the configuration where only the third sub-pattern (enoughBalance) is preserved, and only the corresponding nodes are aligned with expert semantics.

The corresponding results are presented in Figure~\ref{fig:expert_ablation_subpattern}, which contains three 3D bar charts, each representing the classification performance (Accuracy, Recall, Precision, and F1-score) under different sub-pattern configurations for one vulnerability type. A consistent trend is observed: the exclusion of any single expert sub-pattern generally results in a performance drop across all metrics, highlighting the complementary role of these patterns in guiding the model's representation learning. Notably, in the reentrancy and timestamp dependence settings, removing their first sub-pattern (w/o P1) causes the most significant degradation in performance. A plausible explanation is that both P2 and P3 are defined conditionally on the existence of P1. Therefore, when P1 is absent, the higher-level patterns cannot be instantiated, reducing the expert-graph alignment to a trivial form and forcing the model to rely solely on the global graph-level loss, which lacks fine-grained supervision.

Interestingly, we also observe that the exclusion of the third sub-pattern (w/o P3) for results in the smallest impact on model performance across all three vulnerabilities. From a practical standpoint, this can be attributed to the more stringent semantic conditions defined by P3, which leads to fewer graph nodes being labeled. Consequently, the influence of P3 on the local alignment mechanism is inherently limited, rendering its removal less detrimental compared to P1 and P2. Collectively, these findings validate the efficacy of incorporating expert-guided, hierarchical alignment in the distillation process. The dual-focus strategy ensures semantic preservation at both macro and micro levels, while the vulnerability-specific sub-patterns facilitate accurate alignment between the model's latent representations and domain knowledge. 

\section{Conclusions}
In this work, we addressed a chained failure in cross-modal distillation for smart-contract vulnerability detection: because vulnerability is a property of specific nodes rather than the contract as a whole, aligning source and bytecode only at the graph level leaves the student with global structure but without the localized signal that decides a contract's safety. We broke this chain through three coordinated mechanisms. We introduced cross-modal vulnerability patterns that mark corresponding nodes in both modalities, supplying the missing node-level supervision; with these correspondences in place, we developed a dual-attention graph network whose relation-aware attention and adaptive node weighting preserve the vulnerability-critical node signals that uniform encoding would otherwise dilute; given the preserved nodes and the established correspondences, we formulated a dual-focus distillation objective that retains whole-graph alignment while enforcing region-level consistency through its local loss. After distillation the student operates on bytecode alone, requiring no source code at inference---the setting in which most deployed contracts actually reside. On real-world contracts, ExDoS reaches F1 of 90.86\%, 90.23\%, and 83.94\% for reentrancy, timestamp dependency, and infinite loop, improving by 2.7--5.1 points over the strongest per-type baseline. Ablations confirm that each mechanism addresses a distinct link of the failure chain rather than providing redundant gains.

\footnotesize
\bibliographystyle{IEEEtran}
\bibliography{trans-template}

@misc{chainalysis2023euler,
  author       = {Chainalysis},
  title        = {Euler Finance Flash Loan Attack Explained},
  year         = {2023},
  url          = {https://www.chainalysis.com/blog/euler-finance-flash-loan-attack/},

}

@inproceedings{chen2024intro_formal3,
  title={Verifying Declarative Smart Contracts},
  author={Chen, Haoxian and Lu, Lan and Massey, Brendan and Wang, Yuepeng and Loo, Boon Thau},
  booktitle={Proceedings of the IEEE/ACM 46th International Conference on Software Engineering},
  pages={1--12},
  year={2024}
}

@inproceedings{so2021intro_symbolicexe1,
  title={$\{$SmarTest$\}$: Effectively hunting vulnerable transaction sequences in smart contracts through language $\{$Model-Guided$\}$ symbolic execution},
  author={So, Sunbeom and Hong, Seongjoon and Oh, Hakjoo},
  booktitle={30th USENIX Security Symposium (USENIX Security 21)},
  pages={1361--1378},
  year={2021}
}

@inproceedings{yao2022introsymbolicexe2,
  title={An improved vulnerability detection system of smart contracts based on symbolic execution},
  author={Yao, Yao and Li, Hui and Yang, Xin and Le, Yiwang},
  booktitle={2022 IEEE International Conference on Big Data (Big Data)},
  pages={3225--3234},
  year={2022},
  organization={IEEE}
}

@inproceedings{he2019intro_fuzz1,
  title={Learning to fuzz from symbolic execution with application to smart contracts},
  author={He, Jingxuan and Balunovi{\'c}, Mislav and Ambroladze, Nodar and Tsankov, Petar and Vechev, Martin},
  booktitle={Proceedings of the 2019 ACM SIGSAC conference on computer and communications security},
  pages={531--548},
  year={2019}
}

@inproceedings{nguyen2020sfuzz,
  title={sfuzz: An efficient adaptive fuzzer for solidity smart contracts},
  author={Nguyen, Tai D and Pham, Long H and Sun, Jun and Lin, Yun and Minh, Quang Tran},
  booktitle={Proceedings of the ACM/IEEE 42nd International Conference on Software Engineering},
  pages={778--788},
  year={2020}
}

@inproceedings{wu2024handle_expert_challenge,
  title={Are we there yet? unraveling the state-of-the-art smart contract fuzzers},
  author={Wu, Shuohan and Li, Zihao and Yan, Luyi and Chen, Weimin and Jiang, Muhui and Wang, Chenxu and Luo, Xiapu and Zhou, Hao},
  booktitle={Proceedings of the IEEE/ACM 46th International Conference on Software Engineering},
  pages={1--13},
  year={2024}
}

@article{qian2022handle_expert_challenge,
  title={Smart contract vulnerability detection technique: A survey},
  author={Qian, Peng and Liu, Zhenguang and He, Qinming and Huang, Butian and Tian, Duanzheng and Wang, Xun},
  journal={arXiv preprint arXiv:2209.05872},
  year={2022}
}

@article{ding2025LLM4scv_handle_expert_challenge,
  title={SmartGuard: An LLM-enhanced framework for smart contract vulnerability detection},
  author={Ding, Hao and Liu, Yizhou and Piao, Xuefeng and Song, Huihui and Ji, Zhenzhou},
  journal={Expert Systems with Applications},
  volume={269},
  pages={126479},
  year={2025},
  publisher={Elsevier}
}

@article{hu2021dl_lstm,
  title={Transaction-based classification and detection approach for Ethereum smart contract},
  author={Hu, Teng and Liu, Xiaolei and Chen, Ting and Zhang, Xiaosong and Huang, Xiaoming and Niu, Weina and Lu, Jiazhong and Zhou, Kun and Liu, Yuan},
  journal={Information Processing \& Management},
  volume={58},
  number={2},
  pages={102462},
  year={2021},
  publisher={Elsevier}
}

@article{liu2022dl_GraphTransformer,
  title={Blockchain-enabled fraud discovery through abnormal smart contract detection on Ethereum},
  author={Liu, Lin and Tsai, Wei-Tek and Bhuiyan, Md Zakirul Alam and Peng, Hao and Liu, Mingsheng},
  journal={Future Generation Computer Systems},
  volume={128},
  pages={158--166},
  year={2022},
  publisher={Elsevier}
}

@article{chen2021tradition_byteonly,
  title={Defectchecker: Automated smart contract defect detection by analyzing evm bytecode},
  author={Chen, Jiachi and Xia, Xin and Lo, David and Grundy, John and Luo, Xiapu and Chen, Ting},
  journal={IEEE Transactions on Software Engineering},
  volume={48},
  number={7},
  pages={2189--2207},
  year={2021},
  publisher={IEEE}
}

@article{bu2025smartbugbert_byteonly,
  title={Smartbugbert: Bert-enhanced vulnerability detection for smart contract bytecode},
  author={Bu, Jiuyang and Li, Wenkai and Li, Zongwei and Zhang, Zeng and Li, Xiaoqi},
  journal={arXiv preprint arXiv:2504.05002},
  year={2025}
}

@inproceedings{shi2022byteonly,
  title={A bytecode-based approach for smart contract classification},
  author={Shi, Chaochen and Xiang, Yong and Yu, Jiangshan and Gao, Longxiang and Sood, Keshav and Doss, Robin Ram Mohan},
  booktitle={2022 IEEE International Conference on Software Analysis, Evolution and Reengineering (SANER)},
  pages={1046--1054},
  year={2022},
  organization={IEEE}
}

@inproceedings{qian2023cross_www,
  title={Cross-modality mutual learning for enhancing smart contract vulnerability detection on bytecode},
  author={Qian, Peng and Liu, Zhenguang and Yin, Yifang and He, Qinming},
  booktitle={Proceedings of the ACM Web Conference 2023},
  pages={2220--2229},
  year={2023}
}

@inproceedings{sun2025mtvhunter,
  title={MTVHunter: Smart Contracts Vulnerability Detection Based on Multi-Teacher Knowledge Translation},
  author={Sun, Guokai and Zhuang, Yuan and Zhang, Shuo and Feng, Xiaoyu and Liu, Zhenguang and Zhang, Liguo},
  booktitle={Proceedings of the AAAI Conference on Artificial Intelligence},
  volume={39},
  number={14},
  pages={15169--15176},
  year={2025}
}

@article{nakamoto2008bitcoin,
  title={Bitcoin: A peer-to-peer electronic cash system},
  author={Nakamoto, Satoshi},
  year={2008}
}

@article{buterin2013ethereum,
  title={Ethereum white paper},
  author={Buterin, Vitalik and others},
  journal={GitHub repository},
  volume={1},
  number={22-23},
  pages={5--7},
  year={2013}
}

@article{chen2020bytecodeyes_sccodeno_argue,
  title={Defining smart contract defects on ethereum},
  author={Chen, Jiachi and Xia, Xin and Lo, David and Grundy, John and Luo, Xiapu and Chen, Ting},
  journal={IEEE Transactions on Software Engineering},
  volume={48},
  number={1},
  pages={327--345},
  year={2020},
  publisher={IEEE}
}

@article{singh2020bkg2.2_r2,
  title={Blockchain smart contracts formalization: Approaches and challenges to address vulnerabilities},
  author={Singh, Amritraj and Parizi, Reza M and Zhang, Qi and Choo, Kim-Kwang Raymond and Dehghantanha, Ali},
  journal={Computers \& Security},
  volume={88},
  pages={101654},
  year={2020},
  publisher={Elsevier}
}

@inproceedings{mense2018bkg2.2_r3/timestamp,
  title={Security vulnerabilities in ethereum smart contracts},
  author={Mense, Alexander and Flatscher, Markus},
  booktitle={Proceedings of the 20th international conference on information integration and web-based applications \& services},
  pages={375--380},
  year={2018}
}

@article{wu2024bkg2.2_r4_loop,
  title={A Review of Deep Learning-Based Vulnerability Detection Tools for Ethernet Smart Contracts.},
  author={Wu, Huaiguang and Peng, Yibo and He, Yaqiong and Fan, Jinlin},
  journal={CMES-Computer Modeling in Engineering \& Sciences},
  volume={140},
  number={1},
  year={2024}
}

@inproceedings{chen2025clep,
  title={CLEP: A Novel Contrastive Learning Method for Evolutionary Reentrancy Vulnerability Detection},
  author={Chen, Jie and Wang, Liangmin and Zhu, Huijuan and Sheng, Victor S},
  booktitle={Proceedings of the AAAI Conference on Artificial Intelligence},
  volume={39},
  number={1},
  pages={67--74},
  year={2025}
}

@inproceedings{zhao2017dao,
  title={The DAO attack paradoxes in propositional logic},
  author={Zhao, Xiangfu and Chen, Zhongyu and Chen, Xin and Wang, Yanxia and Tang, Changbing},
  booktitle={2017 4th international conference on systems and informatics (ICSAI)},
  pages={1743--1746},
  year={2017},
  organization={IEEE}
}

@inproceedings{tsankov2018securify,
  title={Securify: Practical security analysis of smart contracts},
  author={Tsankov, Petar and Dan, Andrei and Drachsler-Cohen, Dana and Gervais, Arthur and Buenzli, Florian and Vechev, Martin},
  booktitle={Proceedings of the 2018 ACM SIGSAC conference on computer and communications security},
  pages={67--82},
  year={2018}
}

@inproceedings{krupp2018teether,
  title={$\{$teEther$\}$: Gnawing at ethereum to automatically exploit smart contracts},
  author={Krupp, Johannes and Rossow, Christian},
  booktitle={27th USENIX security symposium (USENIX Security 18)},
  pages={1317--1333},
  year={2018}
}

@inproceedings{luu2016making,
  title={Making smart contracts smarter},
  author={Luu, Loi and Chu, Duc-Hiep and Olickel, Hrishi and Saxena, Prateek and Hobor, Aquinas},
  booktitle={Proceedings of the 2016 ACM SIGSAC conference on computer and communications security},
  pages={254--269},
  year={2016}
}

@misc{mueller2017framework,
  author = {Bernhard Mueller},
  title = {A Framework for Bug Hunting on the Ethereum Blockchain},
  howpublished = {\url{https://github.com/ConsenSys/mythril}},
  year = {2017},
}

@article{liu2023rethinking,
  title={Rethinking smart contract fuzzing: Fuzzing with invocation ordering and important branch revisiting},
  author={Liu, Zhenguang and Qian, Peng and Yang, Jiaxu and Liu, Lingfeng and Xu, Xiaojun and He, Qinming and Zhang, Xiaosong},
  journal={IEEE Transactions on Information Forensics and Security},
  volume={18},
  pages={1237--1251},
  year={2023},
  publisher={IEEE}
}

@inproceedings{momeni2019machine,
  title={Machine learning model for smart contracts security analysis},
  author={Momeni, Pouyan and Wang, Yu and Samavi, Reza},
  booktitle={2019 17th international conference on privacy, security and trust (PST)},
  pages={1--6},
  year={2019},
  organization={IEEE}
}

@article{tann2018towards,
  title={Towards safer smart contracts: A sequence learning approach to detecting security threats},
  author={Tann, Wesley Joon-Wie and Han, Xing Jie and Gupta, Sourav Sen and Ong, Yew-Soon},
  journal={arXiv preprint arXiv:1811.06632},
  year={2018}
}

@inproceedings{zhuang2021smart_tmp,
  title={Smart contract vulnerability detection using graph neural networks},
  author={Zhuang, Yuan and Liu, Zhenguang and Qian, Peng and Liu, Qi and Wang, Xiang and He, Qinming},
  booktitle={Proceedings of the Twenty-Ninth International Conference on International Joint Conferences on Artificial Intelligence},
  pages={3283--3290},
  year={2021}
}

@inproceedings{luo2024scvhunter,
  title={Scvhunter: Smart contract vulnerability detection based on heterogeneous graph attention network},
  author={Luo, Feng and Luo, Ruijie and Chen, Ting and Qiao, Ao and He, Zheyuan and Song, Shuwei and Jiang, Yu and Li, Sixing},
  booktitle={Proceedings of the IEEE/ACM 46th International Conference on Software Engineering},
  pages={1--13},
  year={2024}
}

@article{liu2021interpretable_graph_expert_ame,
  title={Smart contract vulnerability detection: from pure neural network to interpretable graph feature and expert pattern fusion},
  author={Liu, Zhenguang and Qian, Peng and Wang, Xiang and Zhu, Lei and He, Qinming and Ji, Shouling},
  journal={arXiv preprint arXiv:2106.09282},
  year={2021}
}

@article{liu2021combining_expert_gnn,
  title={Combining graph neural networks with expert knowledge for smart contract vulnerability detection},
  author={Liu, Zhenguang and Qian, Peng and Wang, Xiaoyang and Zhuang, Yuan and Qiu, Lin and Wang, Xun},
  journal={IEEE Transactions on Knowledge and Data Engineering},
  volume={35},
  number={2},
  pages={1296--1310},
  year={2021},
  publisher={IEEE}
}

@inproceedings{chen2024contrastive_ICSE,
  title={Improving smart contract security with contrastive learning-based vulnerability detection},
  author={Chen, Yizhou and Sun, Zeyu and Gong, Zhihao and Hao, Dan},
  booktitle={Proceedings of the IEEE/ACM 46th International Conference on Software Engineering},
  pages={1--11},
  year={2024}
}

@inproceedings{jiang2018contractfuzzer,
  title={Contractfuzzer: Fuzzing smart contracts for vulnerability detection},
  author={Jiang, Bo and Liu, Ye and Chan, Wing Kwong},
  booktitle={Proceedings of the 33rd ACM/IEEE international conference on automated software engineering},
  pages={259--269},
  year={2018}
}

@inproceedings{kalra2018zeus,
  title={Zeus: analyzing safety of smart contracts.},
  author={Kalra, Sukrit and Goel, Seep and Dhawan, Mohan and Sharma, Subodh},
  booktitle={Ndss},
  pages={1--12},
  year={2018}
}

@inproceedings{wang2023multi_with_miss,
  title={Multi-modal learning with missing modality via shared-specific feature modelling},
  author={Wang, Hu and Chen, Yuanhong and Ma, Congbo and Avery, Jodie and Hull, Louise and Carneiro, Gustavo},
  booktitle={Proceedings of the IEEE/CVF Conference on Computer Vision and Pattern Recognition},
  pages={15878--15887},
  year={2023}
}

@inproceedings{ji2021KD_prior,
  title={Refine myself by teaching myself: Feature refinement via self-knowledge distillation},
  author={Ji, Mingi and Shin, Seungjae and Hwang, Seunghyun and Park, Gibeom and Moon, Il-Chul},
  booktitle={Proceedings of the IEEE/CVF conference on computer vision and pattern recognition},
  pages={10664--10673},
  year={2021}
}

@inproceedings{yang2023KD_prior,
  title={Learning to distill graph neural networks},
  author={Yang, Cheng and Guo, Yuxin and Xu, Yao and Shi, Chuan and Liu, Jiawei and Wang, Chunchen and Li, Xin and Guo, Ning and Yin, Hongzhi},
  booktitle={Proceedings of the sixteenth ACM international conference on web search and data mining},
  pages={123--131},
  year={2023}
}

@inproceedings{tikhomirov2018smartcheck,
  title={Smartcheck: Static analysis of ethereum smart contracts},
  author={Tikhomirov, Sergei and Voskresenskaya, Ekaterina and Ivanitskiy, Ivan and Takhaviev, Ramil and Marchenko, Evgeny and Alexandrov, Yaroslav},
  booktitle={Proceedings of the 1st international workshop on emerging trends in software engineering for blockchain},
  pages={9--16},
  year={2018}
}

@inproceedings{feist2019slither,
  title={Slither: a static analysis framework for smart contracts},
  author={Feist, Josselin and Grieco, Gustavo and Groce, Alex},
  booktitle={2019 IEEE/ACM 2nd International Workshop on Emerging Trends in Software Engineering for Blockchain (WETSEB)},
  pages={8--15},
  year={2019},
  organization={IEEE}
}

@inproceedings{ibing2015pda,
  title={A fixed-point algorithm for automated static detection of infinite loops},
  author={Ibing, Andreas and Mai, Alexandra},
  booktitle={2015 IEEE 16th International Symposium on High Assurance Systems Engineering},
  pages={44--51},
  year={2015},
  organization={IEEE}
}

@inproceedings{burnim2009looper,
  title={Looper: Lightweight detection of infinite loops at runtime},
  author={Burnim, Jacob and Jalbert, Nicholas and Stergiou, Christos and Sen, Koushik},
  booktitle={2009 IEEE/ACM International Conference on Automated Software Engineering},
  pages={161--169},
  year={2009},
  organization={IEEE}
}

@article{tann2018vanilla-rnn,
  title={Towards safer smart contracts: A sequence learning approach to detecting security threats},
  author={Tann, Wesley Joon-Wie and Han, Xing Jie and Gupta, Sourav Sen and Ong, Yew-Soon},
  journal={arXiv preprint arXiv:1811.06632},
  year={2018}
}

@inproceedings{sak2014lstm,
  title={Long short-term memory recurrent neural network architectures for large scale acoustic modeling.},
  author={Sak, Hasim and Senior, Andrew W and Beaufays, Fran{\c{c}}oise and others},
  booktitle={Interspeech},
  volume={2014},
  pages={338--342},
  year={2014}
}

@article{chung2014gru,
  title={Empirical evaluation of gated recurrent neural networks on sequence modeling},
  author={Chung, Junyoung and Gulcehre, Caglar and Cho, KyungHyun and Bengio, Yoshua},
  journal={arXiv preprint arXiv:1412.3555},
  year={2014}
}

@article{kipf2016gcn,
  title={Semi-supervised classification with graph convolutional networks},
  author={Kipf, Thomas N and Welling, Max},
  journal={arXiv preprint arXiv:1609.02907},
  year={2016}
}

@article{vaswani2017attention,
  title={Attention is all you need},
  author={Vaswani, Ashish and Shazeer, Noam and Parmar, Niki and Uszkoreit, Jakob and Jones, Llion and Gomez, Aidan N and Kaiser, {\L}ukasz and Polosukhin, Illia},
  journal={Advances in neural information processing systems},
  volume={30},
  year={2017}
}

@article{duan2023new,
  title={A new smart contract anomaly detection method by fusing opcode and source code features for blockchain services},
  author={Duan, Li and Yang, Liu and Liu, Chunhong and Ni, Wei and Wang, Wei},
  journal={IEEE Transactions on Network and Service Management},
  volume={20},
  number={4},
  pages={4354--4368},
  year={2023},
  publisher={IEEE}
}

@article{chen2025numscout,
  title={NumScout: Unveiling Numerical Defects in Smart Contracts using LLM-Pruning Symbolic Execution},
  author={Chen, Jiachi and Shao, Zhenzhe and Yang, Shuo and Shen, Yiming and Wang, Yanlin and Chen, Ting and Shan, Zhenyu and Zheng, Zibin},
  journal={IEEE Transactions on Software Engineering},
  year={2025},
  publisher={IEEE}
}

@article{ye2024funfuzz,
  title={FunFuzz: A Function-Oriented Fuzzer for Smart Contract Vulnerability Detection with High Effectiveness and Efficiency},
  author={Ye, Mingxi and Nan, Yuhong and Dai, Hong-Ning and Yang, Shuo and Luo, Xiapu and Zheng, Zibin},
  journal={ACM Transactions on Software Engineering and Methodology},
  volume={33},
  number={7},
  pages={1--20},
  year={2024},
  publisher={ACM New York, NY}
}

@article{lin2024crpwarner,
  title={Crpwarner: Warning the risk of contract-related rug pull in defi smart contracts},
  author={Lin, Zewei and Chen, Jiachi and Wu, Jiajing and Zhang, Weizhe and Wang, Yongjuan and Zheng, Zibin},
  journal={IEEE Transactions on Software Engineering},
  volume={50},
  number={6},
  pages={1534--1547},
  year={2024},
  publisher={IEEE}
}

@article{guo2024smart,
  title={Smart contract code repair recommendation based on reinforcement learning and multi-metric optimization},
  author={Guo, Hanyang and Chen, Yingye and Chen, Xiangping and Huang, Yuan and Zheng, Zibin},
  journal={ACM Transactions on Software Engineering and Methodology},
  volume={33},
  number={4},
  pages={1--31},
  year={2024},
  publisher={ACM New York, NY}
}

@article{li2023vulhunter,
  title={Vulhunter: Hunting vulnerable smart contracts at evm bytecode-level via multiple instance learning},
  author={Li, Zhaoxuan and Lu, Siqi and Zhang, Rui and Zhao, Ziming and Liang, Rujin and Xue, Rui and Li, Wenhao and Zhang, Fan and Gao, Sheng},
  journal={IEEE Transactions on Software Engineering},
  volume={49},
  number={11},
  pages={4886--4916},
  year={2023},
  publisher={IEEE}
}

@inproceedings{li2024cobra,
  title={Cobra: interaction-aware bytecode-level vulnerability detector for smart contracts},
  author={Li, Wenkai and Li, Xiaoqi and Li, Zongwei and Zhang, Yuqing},
  booktitle={Proceedings of the 39th IEEE/ACM international conference on automated software engineering},
  pages={1358--1369},
  year={2024}
}

@article{zheng2024dappscan,
  title={Dappscan: building large-scale datasets for smart contract weaknesses in dapp projects},
  author={Zheng, Zibin and Su, Jianzhong and Chen, Jiachi and Lo, David and Zhong, Zhijie and Ye, Mingxi},
  journal={IEEE Transactions on Software Engineering},
  volume={50},
  number={6},
  pages={1360--1373},
  year={2024},
  publisher={IEEE}
}

@article{chen2025chatgpt,
  title={When chatgpt meets smart contract vulnerability detection: How far are we?},
  author={Chen, Chong and Su, Jianzhong and Chen, Jiachi and Wang, Yanlin and Bi, Tingting and Yu, Jianxing and Wang, Yanli and Lin, Xingwei and Chen, Ting and Zheng, Zibin},
  journal={ACM Transactions on Software Engineering and Methodology},
  volume={34},
  number={4},
  pages={1--30},
  year={2025},
  publisher={ACM New York, NY}
}
\end{sloppypar}
\end{document}